\documentclass[twocolumn,twocolappendix]{aastex7}

\usepackage[colorinlistoftodos]{todonotes}
\usepackage{amssymb}
\usepackage{comment}
\usepackage{multirow,acronym}
\usepackage{natbib}
\usepackage{listings}
\usepackage{makecell}
\usepackage{booktabs}
\usepackage{amsmath} 
\usepackage{subfigure}
\usepackage{color}
\usepackage{xcolor}
\usepackage{hyperref}
\usepackage[T1]{fontenc}
\usepackage{graphicx}
\usepackage{bm}
\usepackage{iftex}
\ifXeTeX
  \usepackage{xeCJK}
  \newcommand{\chiname}{(王力乐)}
\else
  \newcommand{\chiname}{}
\fi
\hypersetup{colorlinks=true, citecolor=blue, 
  linkcolor=cyan, urlcolor=magenta}

\usepackage{aas_macros}
\usepackage{float}
\usepackage{amsmath,amssymb}
\usepackage{natbib} % For use with bibtex
\usepackage{graphicx} % For included graphics
\usepackage{color}
\usepackage{verbatim}
\usepackage{rotating}

\newcommand{\proptosim}{\mathrel{\vcenter{
 \offinterlineskip\halign{\hfil$##$\cr
 \propto\cr\noalign{\kern2pt}\sim\cr\noalign{\kern-2pt}}}}}
\renewcommand{\b}[1]{\boldsymbol{#1}}
\renewcommand{\i}{\ensuremath{\rm i}}
\newcommand{\unit}[1]{{\rm\, #1}}

\newcommand{\mean}[1]{\langle #1\rangle}
\renewcommand{\min}{\mathrm{min}}
 
\newcommand{\cm}{\unit{cm}}
\newcommand{\m}{\unit{m}}
\newcommand{\g}{\unit{g}}
\newcommand{\G}{\unit{G}}
\newcommand{\K}{\unit{K}} 
\newcommand{\km}{\unit{km}}

\newcommand{\msun}{M_\odot}

\newcommand{\kpc}{\unit{kpc}}
\newcommand{\Mpc}{\unit{Mpc}}

\newcommand{\keV}{\unit{keV}}
\newcommand{\s}{\mathrm{s}}

\newcommand{\lya}{\text{Ly}\ensuremath{\alpha}}

\newcommand{\B}{\mathbf{B}}     % Magnetic fields
\newcommand{\E}{\mathbf{E}}     % Electric fields
\newcommand{\J}{\mathbf{J}}     % Electric current fields
\newcommand{\F}{\mathbf{F}}     % Fluxes
\renewcommand{\v}{\mathbf{v}}   % Velocity fields
\renewcommand{\d}{\mathrm{d}}
\newcommand{\e}{\mathrm{e}}

\renewcommand{\ion}[2]
{{\rm#1}\;\textsc{\MakeLowercase{#2}}}
\newcommand{\p}{\mathrm{p}}     % Poloidal
\newcommand{\tot}{\mathrm{tot}}
\renewcommand{\sc}{\ensuremath{\mathrm{sc}}}

\renewcommand{\O}{\mathrm{O}}     % Ohmic
\newcommand{\A}{\mathrm{A}}     % Ambipolar
\renewcommand{\H}{\mathrm{H}}     % Hall
\renewcommand{\P}{\mathrm{P}}     % P
\newcommand*\chem[1]{\ensuremath{\mathrm{#1}}}
\renewcommand{\i}{\ensuremath{\mathrm{i}}}
\newcommand{\code}[1]{\lstinline{#1}}
\newcommand{\kratos}{{\lstinline{Kratos}}}
\newcommand{\figdir}{.}

\begin{document}

\title{High Resolution Grid-based Simulations of the Warm-Hot
  Intergalactic Medium}
 
\author[0000-0002-6540-7042]{Lile Wang \chiname}
\affil{The Kavli Institute for Astronomy and Astrophysics,
  Peking University, Beijing 100871, China}
\affil{Department of Astronomy, School of Physics, Peking
  University, Beijing 100871, China}
\email{lilew@pku.edu.cn}

\author[0000-0001-8531-9536]{Renyue Cen}
\affiliation{Institute for Advanced Study in Physics,
  Zhejiang University, Hangzhou 310027, China}
\affiliation{Institute of Astronomy, School of Physics,
  Zhejiang University, Hangzhou 310027, China}
\email{renyuecen@zju.edu.cn}

\correspondingauthor{Lile Wang}
\email{lilew@pku.edu.cn}

\correspondingauthor{Renyue Cen}
\email{renyuecen@zju.edu.cn}

\begin{abstract}
  We present high-resolution cosmological hydrodynamic simulations of the Warm-Hot Intergalactic Medium (WHIM) using the GPU-optimized grid-based hydrodynamic code \kratos{}. Employing a uniform $4096^3$ grid in a $(100~h^{-1}Mpc)^3$ comoving volume, we achieve a spatial resolution of $\sim24.5h^{-1}kpc$, sufficient to resolve the Jeans scale of gas at $T\sim10^4\K$ and $n_H\sim10^{-3}$ to $10^{-2}~{\rm cm}^{-3}$. We find that $\sim23.4\%$ of cosmic baryons reside in the WHIM phase ($T=10^5-10^7$K) at $z=0$, significantly below the $40-50\%$ found in earlier, lower-resolution simulations. Through a suite of lower-resolution simulations, we demonstrate that spatial resolution plays a pivotal role in determining the WHIM fraction: resolving gas near its Jeans scale allows it to reach higher densities, where enhanced radiative cooling transfers a substantial fraction of baryons out of the WHIM temperature range. The remaining WHIM resides predominantly in filaments and accretion-shock structures in the vicinity of halos, where hierarchical structure formation and ongoing hydrodynamic accretion provide continued shock heating. Synthetic observations of \lya{}, \ion{O}{vi}, \ion{O}{vii}, and \ion{O}{viii} emission reveal distinct morphological and kinematic signatures, with \ion{O}{vi} tracing filament-halo interfaces and the X-ray lines probing hotter gas associated with massive halos. These predictions underscore the importance of current and future missions such as XRISM, ATHENA, and HUBS for mapping WHIM thermodynamics and kinematics. 
\end{abstract}


\section{Introduction}
\label{sec:intro}

The Warm-Hot Intergalactic Medium (WHIM) is a crucial
component of the observable universe, characterized by
temperatures ranging from $10^5~\K$ to $10^7~\K$
\citep[e.g.][] {1999ApJ...514....1C,
  2001ApJ...552..473D}. It is predominantly found in the
extensive filamentary structures that interconnect galaxy
clusters, playing a crucial role in the cosmological baryon
census and potentially harboring a substantial fraction of
the ``missing baryons'' in the universe, which are not
accounted for in observable structures such as galaxies and
clusters, or other phases of intergalactic medium (IGM) that
are detected in the ultraviolet (UV) band \citep[e.g.][]
{2003Natur.421..719N, 2008ApJ...679..194D,
 2008ApJS..177...39T, 2012ApJ...759...23S,
 2012ApJ...759..112T}.

While the WHIM is primarily formed by gravitational shock heating, WHIM serves as a
reservoir for baryonic matter and embeds galaxies in the filaments. Consequently, it has a significant impact on the formation of galaxies in the filaments. Its thermal, chemical and dynamical state may also be altered by 
feedback mechanisms from galaxies, such as energy feedback
from supernovae and supermassive black holes,
\citep{2006ApJ...650..560C, 2011ApJ...731...11C}. The WHIM
also imprints a detectable
signature on the Cosmic Microwave Background radiation through
the kinematic Sunyaev-Zel'dovich (kSZ) effect, providing a
probe for the distribution of baryonic matter in the
universe \citep{2009MNRAS.400.1868F}.

Observational missions on WHIM were pioneered by
the Hubble Space Telescope \citep[e.g.][]{2008ApJS..177...39T}
in the UV band for the warm component and 
the Chandra X-ray Observatory for the hot component of
WHIM \citep[e.g.][]{2002ApJ...573..157N,
  2005Natur.433..495N}. The X-Ray Imaging and Spectroscopy
Mission (XRISM), equipped with a micro-calorimeter (Resolve)
and an imager (Xtend), is designed to perform detailed
spectroscopic observations of the WHIM, probing its
temperature, density, and chemical composition
\citep{2020arXiv200304962X}. The forthcoming High-Resolution
Universe-wide Spectroscopic Survey (HUBS) mission, with its
eV-level energy resolution in the $0.1-2~\keV$ band and a 1
square degree field-of-view, is ideal for studying the
WHIM's chemical history and morphology
\citep{2023SCPMA..6699513B}. The Advanced Telescope for
High-ENergy Astrophysics (Athena) will further enhance WHIM
studies with its X-ray Integral Field Unit Spectrometer
(XIFU), allowing in-depth examinations of WHIM filaments at
redshifts less than $z = 2$
\citep{2012arXiv1207.2745B}. 
%Although the James Webb SpaceTelescope (JWST) is not specifically designed for WHIM studies, its detailed infrared data capture capacity cancontribute to the broader understanding of galaxy formation and evolution, which is related to WHIM distribution.

Numerical simulations are essential for modeling the WHIM
and unraveling its evolutionary pathways. Simulating the gas
behavior under the influence of gravity at super-cluster
scales, including the accretion shocks that could heat the
IGM to the WHIM phase, are instrumental in forecasting the
WHIM's detectability, synthesizing its emission and
absorption characteristics across the ultraviolet (UV) and
X-ray spectra. 
As the shock heating to form the WHIM occurs on moderate density regions 
in the intergalactic medium, unigrid simulations are
very effective in capturing this process.
The adopted spatial resolution of $\Delta x\approx
24.5~\kpc~h^{-1}$ also satisfies the Jeans criterion (see
\S\ref{sec:result-obv-lya} and
Appendix~\ref{sec:resolution-analyses}) for gas
of $\sim 10^K$ and density of $10^{-3}-10^{-2}cm^{-3}$,
ensuring that
self-gravitating collapse and accretion-shock heating are
adequately resolved, and more critically, allowing gas
that should cool to cool and condense.
As simulation studies have shown that AGN and star
formation feedback have energetically subdominant contribution to the formation of the WHIM \citep[e.g.][] {2006ApJ...650..560C,
  2011ApJ...731...11C}, such feedback processes are also 
intentionally turned {\it off} in this work in order to
isolate the contribution of gravitational shock heating
to the WHIM budget. 
%We note that this simplification, while motivated by the cited lower-resolution studies, may become less valid at higher resolution where cold dense gas---the phase most sensitive to feedback---is better resolved. 
A simplified
density- and redshift-dependent metal-enrichment
prescription (\S\ref{sec:method-metals}) is incorporated to
allow for metal cooling and also to enable the oxygen-line synthesis in
\S\ref{sec:results-obv}; further resolution-convergence
tests are presented in
Appendix~\ref{sec:resolution-analyses}.

This paper is structured as follows. \S\ref{sec:method}
describes the numerical simulation methods adopted in this
paper, as well as verifications of the related modules and
calculations in various circumstances related to the cosmic
structure formation. \S\ref{sec:results} discusses the
fiducial high resolution simulation on the physical
properties and the morphologies of the WHIM, and
\S\ref{sec:results-obv} elaborates the expected
correspondence from the cosmology models to the expected
physics to be probed by the space telescope projects via
synthetic observations, especially the profiles of related
emission features. \S\ref{sec:summary} summarizes the paper
and expects future extensions to the simulations in this
paper.  

\section{Methods}
\label{sec:method}

\subsection{Numerical Solvers for the WHIM Dynamics}
\label{sec:method-solver}

The WHIM simulations conducted in this paper utilize the
GPU-optimized high performance code \kratos{}
\citep{2025ApJS..277...63W, 2025arXiv250404941W}, which
involves three fundamental modules for the dynamics: (1)
grid-based hydrodynamic module for the baryonic gas (with
PLM reconstruction and HLLC Riemann solver), (2) particle
module for the dark matter (with second-order leap-frog
integrator), and (3) the Poisson solver for the
gravitational potential field. Because all dynamic processes
take place on an cosmic background, expansion of the
Universe (characterized by the scaling factor $a$ and its
time derivative $\dot{a}$) has to be correctly accounted
for. We adopt the transform named ``supercomoving
variables'' \citep{1998MNRAS.297..467M},
\begin{equation}
  \label{eq:sc-transform}
  \begin{split}
    & x \equiv \dfrac{r}{a}\ ,\
      \d t_\sc \equiv \dfrac{\d t}{a^2}\ , \ 
      v_\sc \equiv a u - \dot{a} r\ ,\ 
    \rho_\sc \equiv a^3 \rho\ ,\\
    & \phi_\sc \equiv a^2
      \left(\phi + \dfrac{a \ddot{a}x^2}{2}\right )\ ,\ 
      p_\sc \equiv a^5 p\ ,\ \epsilon_\sc \equiv a^5 \epsilon\ ,
  \end{split}
\end{equation}
where $r$ is the physical coordinates, $x$ is the comoving
coordinates, $v$ and $u$ are the peculiar and overall
velocity vectors respecitvely, $p$ is the gas pressure,
$\epsilon$ is the total energy density (the sum of internal
and kinetic energy density), and $\rho$ is the mass density.
The variables with subscripts ``sc'' denote the
``supercomoving variables'', the equivalent physical
quantities in the supercomoving system.  Under the
supercomoving transform, the hydrodynamic equations for the
supercomoving variables become (the $\nabla_x$ operator also
denotes the spatial derivatives with respect to the comoving
coordinate $x$; $I$ is the identity tensor),
\begin{equation}
  \label{eq:sc-hydro}
  \begin{split}
    & \dfrac{\partial \rho_\sc}{\partial t_\sc} + \nabla_x
      \cdot (\rho_\sc v_\sc) = 0\ ,\\
    & \dfrac{\partial \rho_\sc v_\sc}{\partial t_\sc}
      + \nabla_x \cdot (\rho_\sc v_\sc v_\sc + p_\sc I) +
      \rho_\sc \nabla_x \phi_\sc = 0\ ,\\
    & \dfrac{\partial \epsilon_\sc}{\partial t_\sc} + \nabla_x
      \cdot [(\epsilon_\sc + p_\sc) v_\sc] + \rho_\sc v_\sc\cdot
      \nabla_x \phi_\sc \\
    & \quad + \mathcal{H} \epsilon_\sc(3\gamma - 5) =
      \Gamma_\sc \ ,\\ 
  \end{split}
\end{equation}
where $\mathcal{H}\equiv a\dot{a}$ is the Hubble parameter
in the supercomoving frame, $\gamma$ is the adiabatic index
of the gas, and $\Gamma_\sc$ collects the heating and cooling
rates (energy change per volume per unit time): the heating
comprises photoheating by the metagalactic ultraviolet
background of \citet{1996ApJ...461...20H} (updated
calibrations are available in \citealt{2012ApJ...746..125H})
together with shock
and adiabatic heating, while the cooling is evaluated as
described below. For gases that consist of ionized species
and single-atom molecules, $\gamma = 5/3$ yields a vanishing
Hubble term. In practice, the hydrodynamic module of \kratos
solves the supercomiving equations \eqref{eq:sc-hydro}
directly in the comoving frame, while the thermodynamics
represented by $\Gamma$ are evolved by, in every timestep:
\begin{enumerate}
\item Convert the internal energy density from its
  supercomoving value to the physical value;
\item Advance the non-equilibrium ionization and recombination
  of the gas, evaluate the cooling rate from the resulting
  non-equilibrium ionic fractions, and integrate the cooling
  equation in physical variables for each cell over the
  physical timestep (with $\delta t = a^2 \delta t_\sc$) using
  an implicit method, with the atomic rates and thermal
  functions of \citet{2012ApJS..202...13G};
\item Convert the evolved physical energy density back to the
  supercomoving energy density, and write the result back to
  the supercomoving hydrodynamics.
\end{enumerate}

For the particles representing dark matters, the dynamics is
evolved with the particle mesh (PM) method.
At the fiducial resolution $N=4096$, the dark-matter particle
count is $(N/2)^3 = 2048^3 \approx 8.6\times10^9$, and the
gravitational softening length equals the cell size
$\Delta x = L/N$ by construction of the PM method.
The CFL safety factor is set to $0.3$.  In the
supercomoving frame, the particle motion is simply described
by $\d v_\sc/\d t_\sc \equiv -\nabla_x \phi_\sc$, and
integrated with the symplectic leap frog method \citep[e.g.][]
{2002nrca.book.....P}. These particles contribute to the
mass density field by ``depositing'' its own density by the
``cloud-in-cell'' method. The momentum conservation is
guaranteed by using the same window function for depositing
particle mass onto the mass density field, and sampling the
acceleration field for particles \citep[see also, e.g.][]
{2016SAAS...43..251S}.

At each timestep, the fluids (including cooling and heating)
and particles are evolved, and the consequent field of mass
density combining the contribution of fluids and particles
is the source field for the subsequent solver for the
gravitation field, on the Poisson equation in the
supercomoving frame,
\begin{equation}
  \label{eq:sc-poisson}
  \nabla_x^2 \tilde{\phi}_x = 4\pi G (\rho_{x,\tot} -
  \mean{\rho_{x,\tot}})\ ,
\end{equation}
where $\rho_{x,\tot} \equiv \rho_\sc$ is the total
supercomoving mass density combining baryonic and dark matter,
and $\mean{\rho_{x,\tot}}$ its spatial average. The
gravitational potential entering both the fluid and particle
dynamics is the same supercomoving potential $\phi_\sc$ of
Eq.~\eqref{eq:sc-hydro}, recovered from the solved equivalent
potential by $\phi_\sc \equiv a \tilde{\phi}_x$. The
Poisson equation \eqref{eq:sc-poisson} is solved with the
GPU-optimized multi-grid solver within \kratos{}, and the
dynamics of the following step is evaluated based on
$\phi_\sc$.

\begin{figure}
  \centering
  \includegraphics[width=3.2in, keepaspectratio]
  {\figdir/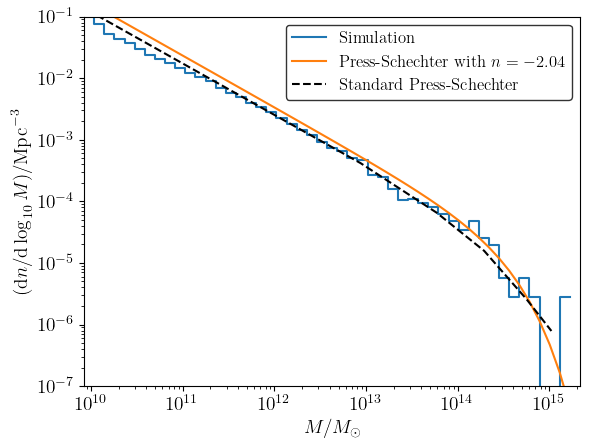}
  \caption{The halo mass function of the fiducial model,
    comparing the \kratos{} simulation results to the
    Press-Schechter mass function \citep{1974ApJ...187..425P}
    assuming index $n = -2.04$ (orange line) and the standard
    Press-Schechter function (dashed line).}
  \label{fig:halo_mass_func} 
\end{figure}

\subsection{The Fiducial Model}
\label{sec:method-fiducial}

\begin{figure*}
  \centering
  \hspace*{-1.5cm}
  \includegraphics[width=7.5in, keepaspectratio]
  {\figdir/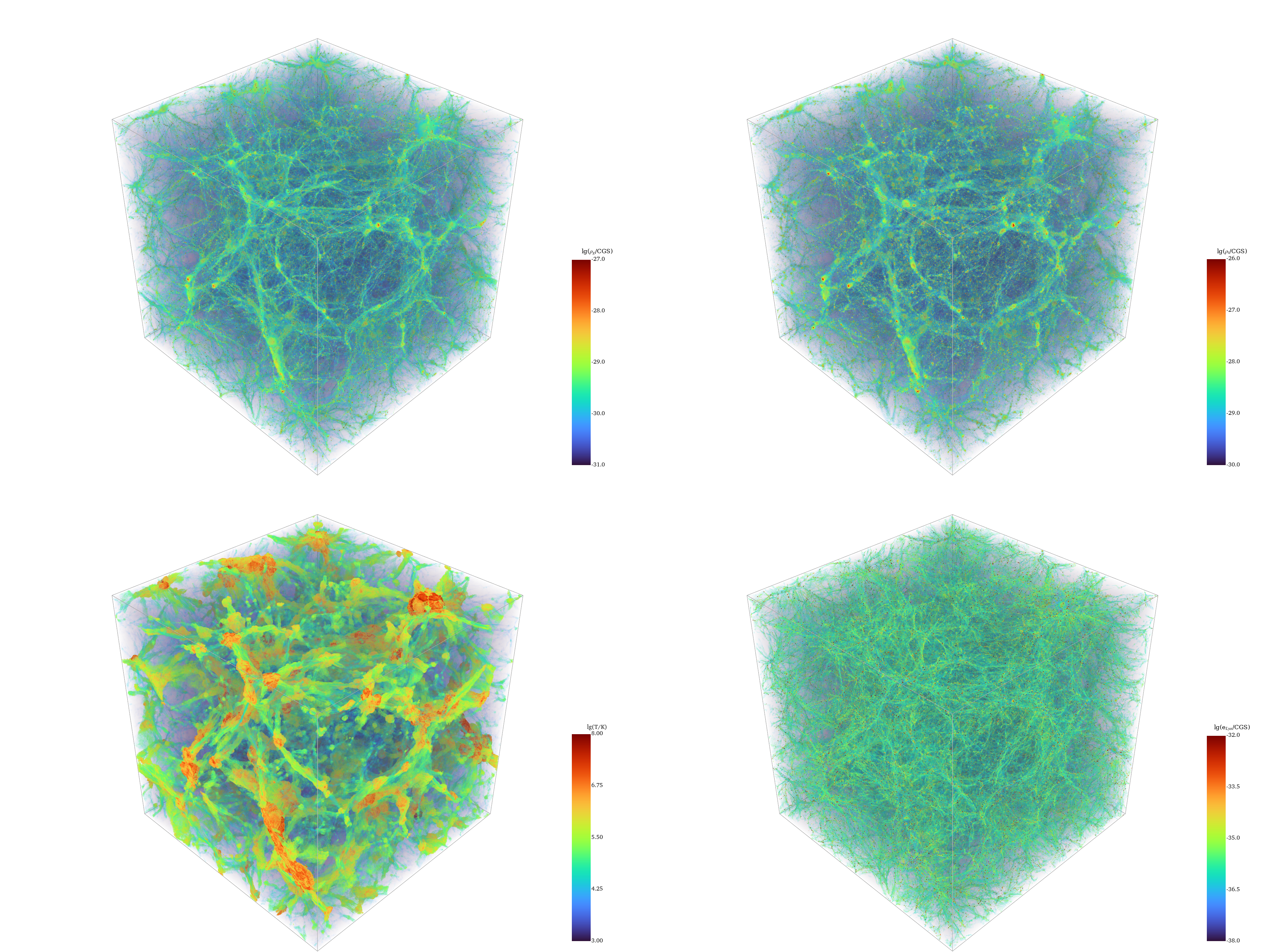}
  \caption{The volume rendering of the key hydrodynamic data
    at $z=0$ for the fiducial cosmology simulation using
    \kratos{} (see also \S\ref{sec:method-fiducial}). The
    upper row presents the three-dimensional morphologies of
    baryonic mass ($\rho_{\rm g}$, upper left) and total mass
    ($\rho_{\rm tot}$ for the combination of baryonic and
    dark matters; upper right). The lower row shows the
    temperatrue (lower left) and \lya{} emissivity (lower
    right; see also \S\ref{sec:result-obv-lya} for details)
    rendering plots. }
  \label{fig:cosmic_render} 
\end{figure*}

To facilitate comparisons with the upcoming observational missions probing the WHIM,
the simulations take the Planck cosmology parameters: $\Omega_k = 0$,
$\Omega_m = 0.315$, $\Omega_\Lambda = 0.685$,
$h \equiv H / (100~\km~\s^{-1}~\Mpc^{-1}) = 0.674$, and
$\sigma_8 = 0.811$ \citep{2020A&A...641A...6P}. The initial
condition is generated by the \code{MUSIC} (Multi-Scale
Initial Conditions) initial condition generator
\citep{2011MNRAS.415.2101H, 2021MNRAS.503..426H} at 
$z = 49$. 
%We have also tested several other reasonablevalues of handover redshift ($z = 19,\ 99$, etc.), and the results do not exhibit any appreciable differences.
 
The simulations are conducted with periodic boundary
conditions. In order to cover a reasonable large spatial
region that contains typical super clusters of galaxies, we
choose the box size $L_{\rm box} = 100~\Mpc~h^{-1}$, with
number of zones $N = 4096$ along each side, so that the
spatial resolution is
$\Delta x = L_{\rm box}/N = 24.5~\kpc~h^{-1}$ (note that all
spatial scales and distances are described with the comoving
frame), 
adequate for resolving the radius of halos more massive than $10^{11}~\msun$ more than 10 cells.  
To address the critical issue of how resolution affects the outcome of WHIM formation and its mass budget at $z=0$,
we also conduct simulations with $N = 512,\ 1024\ ,2048$,
under consistent initial conditions (i.e., initial
conditions for coarser simulations are rebinned from the one
for the fiducial simulation), and the results are compared
to the fiducial $N = 4096$ simulation.

From the technical side, the fiducial simulation poses
challenges on the computational devices. Given the demands
on the computation speed and memory, the fiducial simulation
is carried out on a computer cluster equipped with 512
high-performance GPU accelerators, each with 16~GB of
graphics memory. This simulation
is finished within $\sim 70~{\rm hr}$ wall-clock time. The
consumption of graphics memory (VRAM) on the GPUs is a
major constraint for simulations of this scale: storing the
dark-matter particles at full precision for a $4096^3$
particle load alone would far exceed the 16~GB available on
each device. The \kratos{} system therefore adopts
significant compression for the dark-matter particles,
inspired by the information-aware particle-compression
algorithms developed for the CUBE and CUBE2 codes
\citep{2018ApJS..237...24Y, 2026SCPMA..6969511Y}, which
retain the clustering and other abstract information of the particle
distribution at a small fraction of the raw memory cost
(see \citealt{2018ApJS..237...24Y} and
\citealt{2026SCPMA..6969511Y} for quantitative details of
the compression ratio and error properties).
This compression is essential for enabling the run to fit
within the limited per-device memory.
For reference, the wall-clock times for the lower-resolution
runs are: $N=512$ in $\sim 30$~min on 2$\times$RTX~3090,
$N=1024$ in $\sim 2$~hr on an 8-GPU node with 8$\times$RTX~4090,
and $N=2048$ in $\sim 90$~hr on 16 GPU accelerators with 64~GB
each (a different memory configuration from the fiducial run).

\subsection{Metal Enrichment and Oxygen Abundance}
\label{sec:method-metals}

Because the energy and momentum feedback from galaxy formation
are intentionally switched off in this work
(\S\ref{sec:method-fiducial}), metals are not produced
self-consistently by the simulation. To include metal cooling and also enable the synthetic
oxygen-line observations of \S\ref{sec:results-obv}, we instead
evolve an imposed metallicity field. Each gas cell carries a
metallicity $Z$ advanced by an enrichment rate written as a
derivative with respect to the cosmic scaling factor,
\begin{equation}
  \label{eq:metal-rate}
  \dfrac{\d Z}{\d a} = f(\delta_\rho, a)\ ,\ 
  Z(a) = \min\!\left[\int^{a} f(\delta_\rho, a')\,\d a',\
    0.3 Z_\odot\right]\ ,
\end{equation}
where $\delta_\rho$ is the local baryonic overdensity. The
function $f$ is obtained by numerically interpolating the
chemical-evolution results of \citet{1999ApJ...519L.109C},
who find the metallicity to be a strongly increasing
function of local density at every epoch, rising from
$\sim 1\%~Z_\odot$ globally at $z=3$ to $\sim 20\%~Z_\odot$
by $z=0$ while saturating near solar in the densest
regions. The explicit functional form of $f(\delta_\rho, a)$
and its calibration parameters will be documented in a
forthcoming methods paper (Wang \& Cen, in preparation). The
metallicity is integrated for every cell at each timestep
alongside the hydrodynamics, and capped at $0.3\,Z_\odot$
(defined by a metal mass fraction
$M_{\rm metal}/M_{\rm baryon} = 0.006$) to reproduce this
saturation. Assuming solar relative abundances
[$12 + \log_{10}({\rm O/H})_\odot = 8.69$; \citealt
{2009ARA&A..47..481A}], the oxygen number density follows as
$n_{\rm O} = (Z/Z_\odot)\,({\rm O/H})_\odot\,n_{\rm
  H}$. This abundance enters both the non-equilibrium
cooling evaluated in \S\ref{sec:method-solver} and the
\ion{O}{vi}, \ion{O}{vii}, and \ion{O}{viii} line
emissivities of \S\ref{sec:results-obv}. The prescription
thus captures the density- and redshift-dependent enrichment
that shapes the relative morphologies of the oxygen tracers,
without an explicit model of star formation and feedback.

\section{Results of Simulations}
\label{sec:results}

\begin{figure*}
  \centering
  \hspace*{-0.4cm}
  \includegraphics[width=0.94\textwidth, keepaspectratio]
  {\figdir/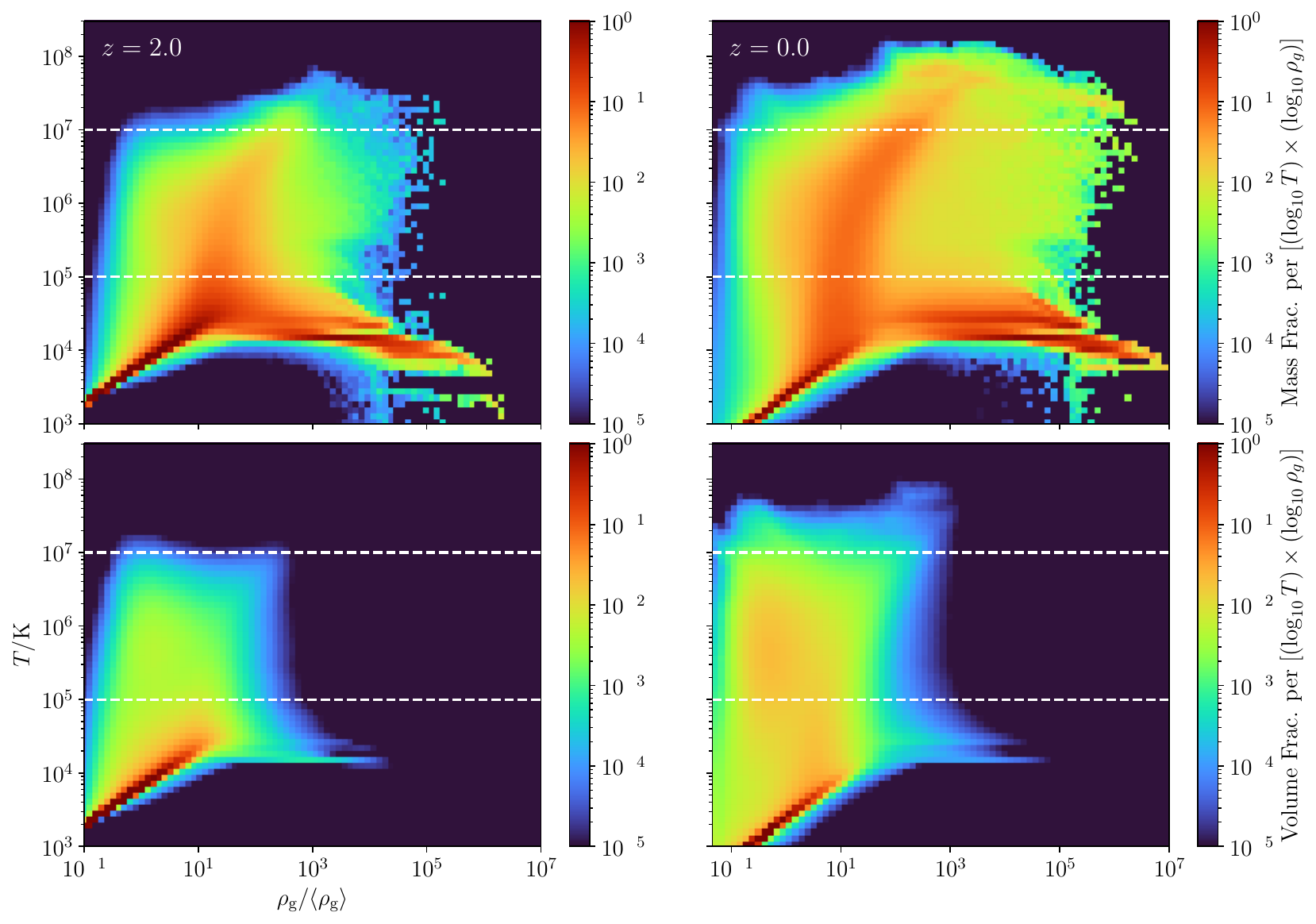}
  \caption{Distribution of gas in the phase space spanned by
    temperature $T$ and comoving gas density $\rho_{\rm g}$
    for the mass-weighted (upper row) and volume-weighted
    (lower row) distributions, showing the snapshots at
    cosmic scaling factors $a = 0.3$ ($z=2.33$; left row)
    and $a = 1$ ($z=0$; right row), respectively. The two
    horizontal dashed lines indicates the lower limit
    ($10^5~\K$) and the upper limits ($10^7~\K$) of
    temperature for the WHIM gas. }
  \label{fig:histogram_dual} 
\end{figure*} 

\subsection{Formation of Structures}
\label{sec:results-struct}

A series of verification tests validate the cosmological
simulations within the \kratos{} framework, including
comparisons with analytical predictions and consistency
checks against observational data; resolution-convergence tests are
detailed in Appendix~\ref{sec:resolution-analyses}. One
particularly critical validation is demonstrated in
Figure~\ref{fig:halo_mass_func}, which compares the
simulated halo mass function to the analytic expectation
from the Press-Schechter formalism
\citep{1974ApJ...187..425P}. The close agreement confirms
that the simulation properly resolves the hierarchical
growth of dark matter halos across cosmic time. A more
stringent comparison against the
\citet{2008ApJ...688..709T} mass function, which is accurate
at the $\sim 5\%$ level, will be presented in a companion
study.

At redshift $z=0$, the large-scale structure of the universe
is visualized through volume-rendered projections in
Figure~\ref{fig:cosmic_render}. On supercluster scales
($\gtrsim 10~\Mpc$), both dark matter and baryonic matter
exhibit similar spatial distributions, reflecting the
dominant role of gravitational force there.
%and tidal forces in structuring the universe. This morphological similarityarises because dark matter, which constitutes $\sim 85\%$ of the total matter content, governs the gravitational potential wells in which baryons accumulate. 
However,
thermodynamic processes introduce deviations at smaller
scales ($\lesssim 5~\Mpc$) particularly in the intrahalo and
intergalactic media.

The phase-space distribution in
Figure~\ref{fig:histogram_dual} further contextualizes the
cosmological significance of WHIM. While the WHIM occupies only 
a few percent of the volume
by $z=0$, it contains $\sim 23.4\%$ of all baryons, a
consequence of its overdensity of $10-100$. 
This value is lower than the $40-50\%$ reported by
\citet{1999ApJ...519L.109C} and \citet{2006ApJ...650..560C}.
recent tSZ and X-ray detections of cosmic-web gas
\citep{2019A&A...618A..38D, 2019MNRAS.483..223T,
  2015Sci...348..779E} support a filamentary baryon
reservoir at the $\sim 30\%$ level, although earlier work
has indicated that SPH based simulations tend to under-produce
WHIM gas than grid-based simulations, in part due to differences
in shock capturing in the two kinds of codes 
\citep{1994ApJ...430...83K}. Our resolution-series analysis in
Appendix~\ref{sec:resolution-analyses} suggests that
lower-resolution simulations systematically overestimate
the WHIM fraction.
%though the absence of feedback in our runs may bias the value low, so the net comparison isnon-trivial. 
The quoted fraction is drawn from a single
$(100~\Mpc~h^{-1})^3$ realization; cosmic variance at this
box size is expected at the several-percent level.
Grid-based (Eulerian) hydrodynamics, although
computationally expensive, is well suited to modeling this
diffuse medium because its formulation captures shock fronts
and entropy gradients in the IGM with high fidelity. These
properties motivate our choice of an Eulerian scheme for the
low-density, shock-dominated WHIM, complementing the
strengths and limitations of smoothed-particle hydrodynamics
(SPH) characterized in code-comparison studies
\citep{2007MNRAS.380..963A, 2012MNRAS.424.2999S}.

%and the observational census of \citet{2012ApJ...759...23S};

\subsection{Thermal Evolution of Intergalactic Gas}
\label{sec:results-thermal}

\begin{figure*}
  \centering
  \hspace*{-0.4cm}
  \includegraphics[width=6.5in, keepaspectratio]
  {\figdir/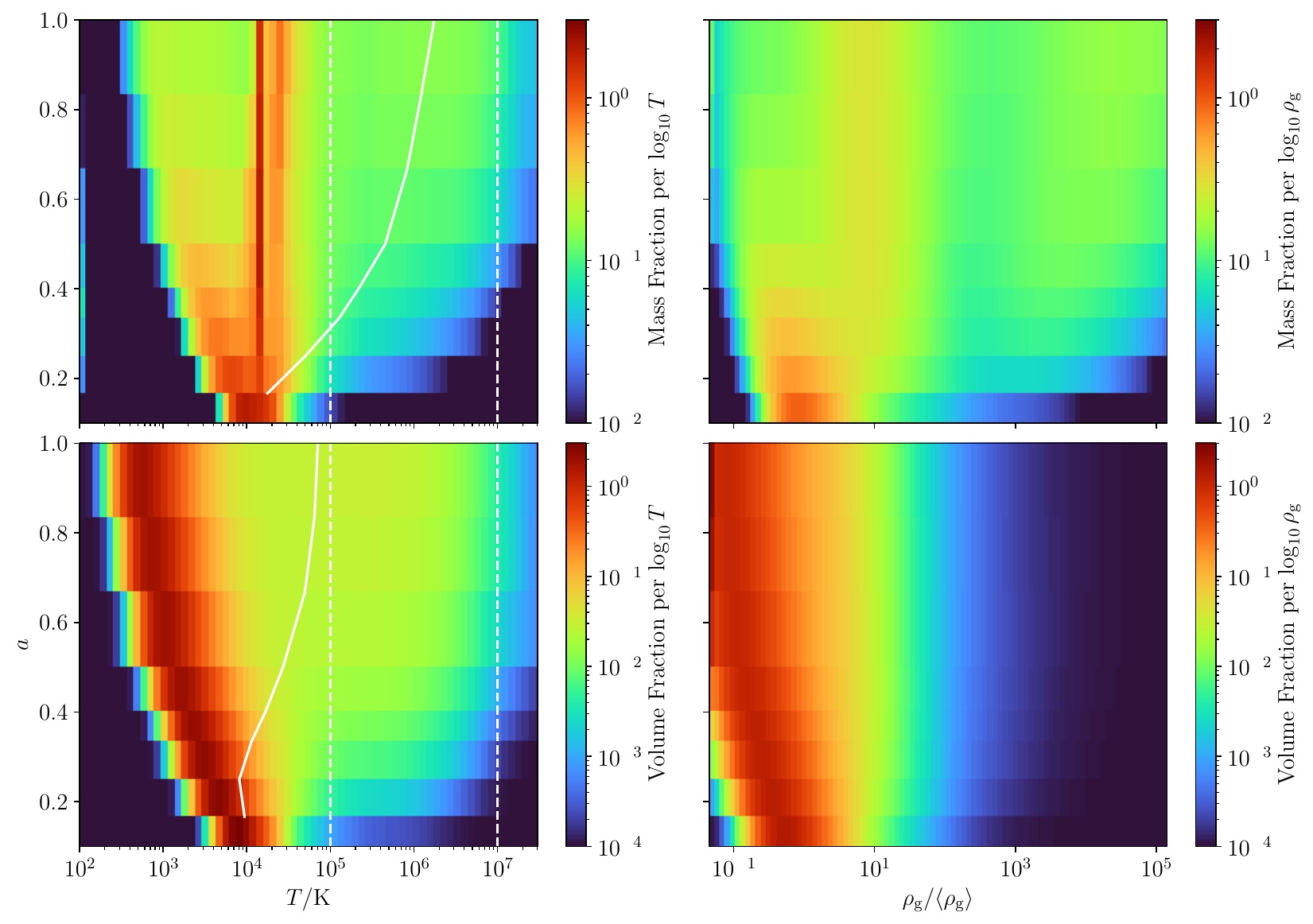}
  \caption{The evolution of mass-weighted (upper row) and
    volume-weighted (lower row) distributions in temperature
    (left column) and comoving gas density (right column),
    indicated by the functions at different cosmic scaling
    factors $a$ (vertical axes). In the left row, the solid
    line indicates the evolution of mass-weighted and
    volume-weighted averages of gas temperature, and the two
    vertical dashed lines mark the temperature boundaries
    for the WHIM gas.}
  \label{fig:histogram_evolve} 
\end{figure*}

The temperature panel in Figure~\ref{fig:cosmic_render}
reveals that the WHIM gas with $T > 10^5~\K$ predominantly
occupies regions surrounding massive galaxy clusters
($\gtrsim 10^{13}~M_\odot$) and the filamentary structures
connecting them. This spatial correlation arises because
accretion shocks generated during structure formation heat
gas to WHIM temperatures. On the shock around halos 
the characteristic virial temperature is,
\begin{equation}
  \label{eq:T_vir}
  T_{\rm vir} \sim 10^5~\K\times
  \left( \dfrac{M_{\rm halo}}{10^{11}~M_\odot} \right)^{2/3}\ ,
\end{equation}
which implies that halos below $\sim 10^{11}~M_\odot$ (e.g.,
dwarf galaxies) cannot sustain virialized gas reservoirs at
WHIM temperatures. Lower dimensional shocks due to formation
of pancakes and filaments are part of the energy sources for the heated
cosmic web.
%Consequently, their circumgalactic media remain predominantly cool ($\sim 10^4~\K$) and are more susceptible to photoionization or galactic winds.

The thermal evolution of baryons provides a critical
diagnostic for understanding cosmic structure formation and
the present-day distribution of ordinary
matter. Figure~\ref{fig:histogram_evolve} quantifies this
evolution through redshift-dependent temperature-density
phase diagrams, revealing two key trends.  The first is a
monotonic rise in volume-weighted mean temperature from
$\langle T \rangle_V \sim 10^2~\K$ at $z=99$ to
$\sim 10^{4.5}~\K$ at $z=0$. In the mean time, a
corresponding increase in mass-weighted mean temperature
from $\langle T \rangle_M \sim 10^5~\K$ at $z=3$ to
$\sim 10^{6}~\K$ at present epoch.

%This heating history directly traces the hierarchical growth
%of dark matter halos and the large-scale structure %\citep{2006ApJ...650..560C}. As
%primordial gas falls into gravitational potential wells,
%kinetic energy from gravitational collapse is thermalized
%through accretion shocks with Mach numbers
%$\mathcal{M} \sim 10-100$, governed by the Rankine-Hugoniot
%jump conditions (assuming $\gamma = 5/3$),
%\begin{equation}
%  \dfrac{T_1}{T_0} \simeq \dfrac
%  {5\mathcal{M}^4 + 14\mathcal{M}^2 - 3}
%  {16\mathcal{M}^2} \ ,
%\end{equation}
%where $T_0$ and $T_1$ are pre- and post-shock temperatures,
%respectively. These shocks efficiently convert dark-matter
%gravitational potential energy into baryonic thermal energy.
%The post-shock temperature rises steeply with the shock Mach
%number, while the resulting virialized gas obeys the virial
%relation $T_{\rm vir} \propto M_{\rm halo}^{2/3}$
%(Eq.~\ref{eq:T_vir}), with density increasing via
%compression.

The phase diagrams in Figure~\ref{fig:histogram_evolve}
trace the evolution of a few different thermal populations
of intergalactic gas: cold condensed gas ($T < 10^4$~K),
warm \lya{} forest gas ($10^4 < T < 10^5$~K), WHIM
($10^5 < T < 10^7$~K), and virialized cluster gas (``hot
gas'', $T > 10^7$~K). The WHIM component emerges as the
dominant baryon reservoir with increasing amount 
since $z\simeq 3$, and grows to
$\sim 23.4\%$ of all baryons by $z=0$. This process occurs
as progressively larger nonlinear structures with higher converging velocities form, with
their associated shock fronts heating previously warm gas to
temperatures where hydrogen becomes fully ionized and
traditional \lya{} forest constitutes a progressively smaller portion of the overall IGM.

\begin{figure}
  \centering
  \includegraphics[width=\linewidth]
  {\figdir/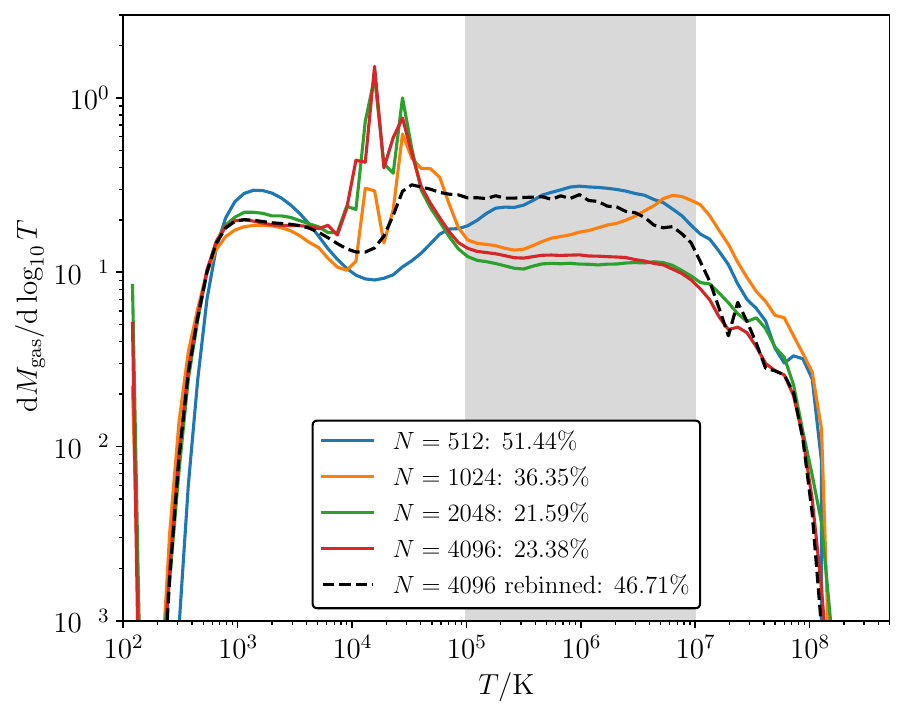}
  \caption{Mass distribution functions over gas temperature,
    ${\rm d}M_{\rm gas}/{\rm d}\log_{10}T$ (normalized to
    unit integral), at $z\simeq 0$ for the resolution series
    $N=512$, 1024, 2048, and 4096, and for the $N=4096$
    output rebinned by a factor of four along each dimension
    (black dashed). The gray band marks the WHIM temperature
    window $10^5$--$10^7~\K$, and the legend quotes the WHIM
    mass fraction of each configuration. Under-resolved
    configurations overproduce gas at WHIM temperatures: the
    fraction drops from $51.4\%$ ($N=512$) and $36.4\%$
    ($N=1024$) to $21.6\%$ ($N=2048$) and $23.4\%$
    ($N=4096$), while rebinning the $N=4096$ output raises
    it back to $46.7\%$. }
  \label{fig:frac-whim}
\end{figure}

The WHIM fraction is, however,
sensitive to numerical resolution. Figure~\ref{fig:frac-whim}
compares the $z\simeq0$ gas mass distribution over temperature
across our resolution series: the WHIM fraction declines from
$51.4\%$ at $N=512$ to $23.4\%$ at the fiducial $N=4096$. There is a big jump down from $1024$ to $2048$, while from $2048$ to $4096$ the change is within noise.
This is a key outcome, indicating that resolving the 
relevant Jeans scale with about $50~h^{-1}~{\rm kpc}$ resolution is imperative to obtain converged results.
Physically, 
resolving the relevant gas allows the cold gas to 
remain cold or to reach
higher densities to then cool out of the WHIM window
(Appendix~\ref{sec:resolution-analyses} presents the full
resolution analysis).
% {\bf RC: If you have a redshift evolution of WHIM fraction comparsions among the runs, it will be very informative.}

\begin{figure*}[ht!]
  \centering
  \hspace*{-0.7cm}
  \includegraphics[width=0.49\textwidth, keepaspectratio]
  {\figdir/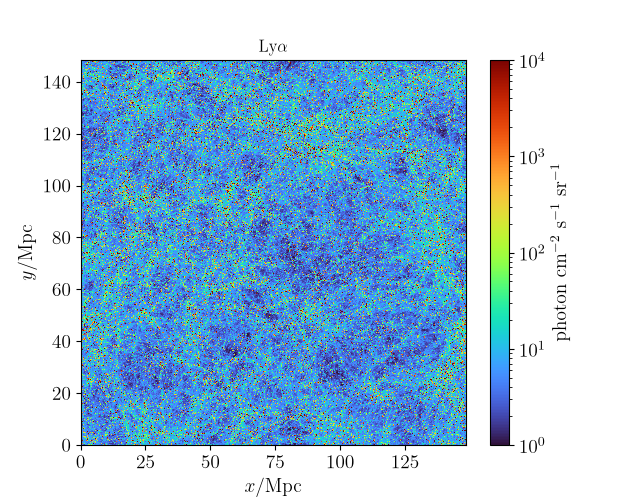}
  \includegraphics[width=0.49\textwidth, keepaspectratio]
  {\figdir/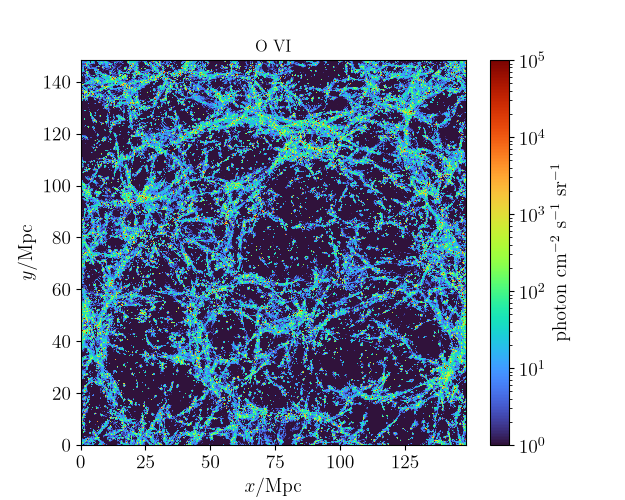} \\
  \includegraphics[width=0.49\textwidth, keepaspectratio]
  {\figdir/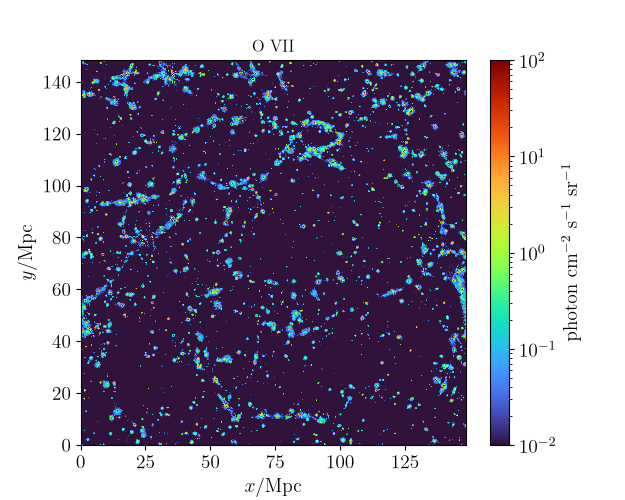}
  \includegraphics[width=0.49\textwidth, keepaspectratio]
  {\figdir/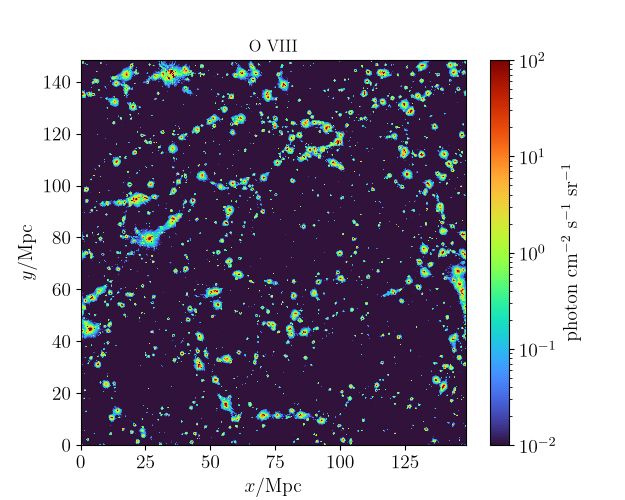}
  \caption{The projected surface brightness over the
    $100~\Mpc~h^{-1}$ simulation domain, for \lya{} (upper left panel),
    \ion{O}{vi} $\lambda\lambda 1032, 1038$~\AA (upper
    right),  \ion{O}{vii} $\lambda 21.6$~\AA (lower left),
    and \ion{O}{viii} $\lambda 19.0$~\AA (lower right)
    emission lines, respectively.
  }
  \label{fig:emis_sb} 
\end{figure*}

\begin{figure}
  \centering
  \includegraphics[width=\columnwidth, keepaspectratio]
  {\figdir/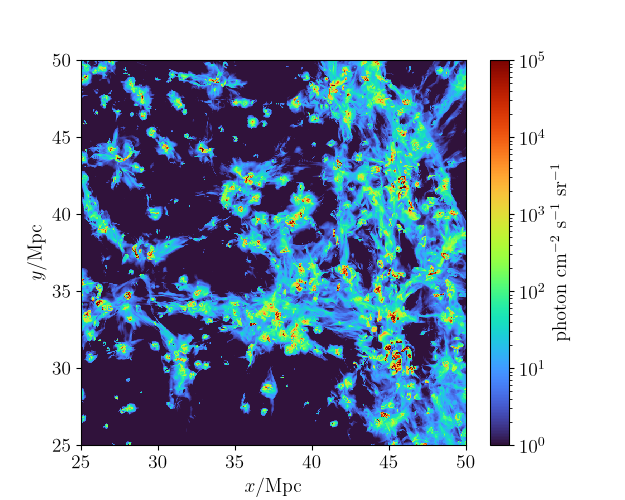}
  \caption{The zoom-in of the \ion{O}{vi} panel in
    Figure~\ref{fig:emis_sb}, focusing on the
    $25~\Mpc~h^{-1}<\{x,y\} < 50~\Mpc~h^{-1}$ region. }
  \label{fig:emis_o6_zoomin} 
\end{figure}

The survival time of WHIM gas against radiative cooling
exceeds the Hubble time for most of this component. The
cooling time at $T\gtrsim 10^6~\K$ temperatures scales
approximately as,
\begin{equation}
  t_{\rm cool} \sim \dfrac{3 n k T}{2 n_e n_{\rm H}\Lambda(T,Z)}
  \propto \dfrac{T^{1/2}}{\delta_0^2(1+z)^3}\ , 
\end{equation}
during the linear growth epoch that has
$\delta \proptosim \delta_0 a^{3/2}$, where $\delta$ is the
dimensionless density contrast and $\Lambda(T,Z)$ the
cooling function, with contributions from bremsstrahlung and
metal-line cooling (the $T^{1/2}$ scaling applies in the
bremsstrahlung-dominated regime, $T\gtrsim 10^7~\K$). For typical
parameters ($\delta = 30$, and taking the $T = 10^6~\K$ from
the mass-weighted expectation value),
$t_{\rm cool} \sim 3t_{\chem{H}}$, allowing the gas to
remain in the warm and hot phases despite metal cooling.
This longevity is further enhanced by ongoing heating from
structure formation shocks and adiabatic compression during
cosmic web assembly.

One can nonetheless identify the peaks in the temperature
distributions near $10^4~\K$. According to the cooling rate
features adopted in this work \citep[see also][]
{2012ApJS..199...20G}, the gas can cool down to $10^4~\K$
only if it has not been heated above $\sim 10^{5.5}~\K$, and will
very likely remain cold once it cools down into blobs that
cannot be easily destroyed by the shock heating processes, 
resulting in the formation of two
relatively distinct ``phases''.  Individual gas elements
undergo multiple phase transitions during cosmic
history. Typical warm/hot gas at $z=0$ has been shock-heated
a few times since $z=1$, spending $\sim 30\%$ of its
time in the $10^5$--$10^7$~K range. These statistics are
estimated from the Eulerian phase fractions (mass in each
temperature bin versus total gas mass) measured at each
snapshot; because the code is Eulerian rather than
Lagrangian, the ``time spent'' figures represent
ensemble-averaged phase occupancy across the resolved gas
population, not tracking of individual fluid elements.
Such heating prevents
permanent incorporation into galaxies, with only $\sim 10\%$
of warm/hot gas eventually cooling into stars by $z=0$
(estimated by $T\lesssim 10^2~\K$). The phase transition
rate, defined as the time derivative of the WHIM mass
fraction, peaks at $z \sim 0.5$, coinciding with the epoch of
maximum WHIM gas fraction.

\section{Synthesis of Observations on WHIM Morphologies}
\label{sec:results-obv}

\begin{figure*}
  \centering
  \hspace*{-0.5cm}
  \includegraphics[width=7.2in, keepaspectratio]
  {\figdir/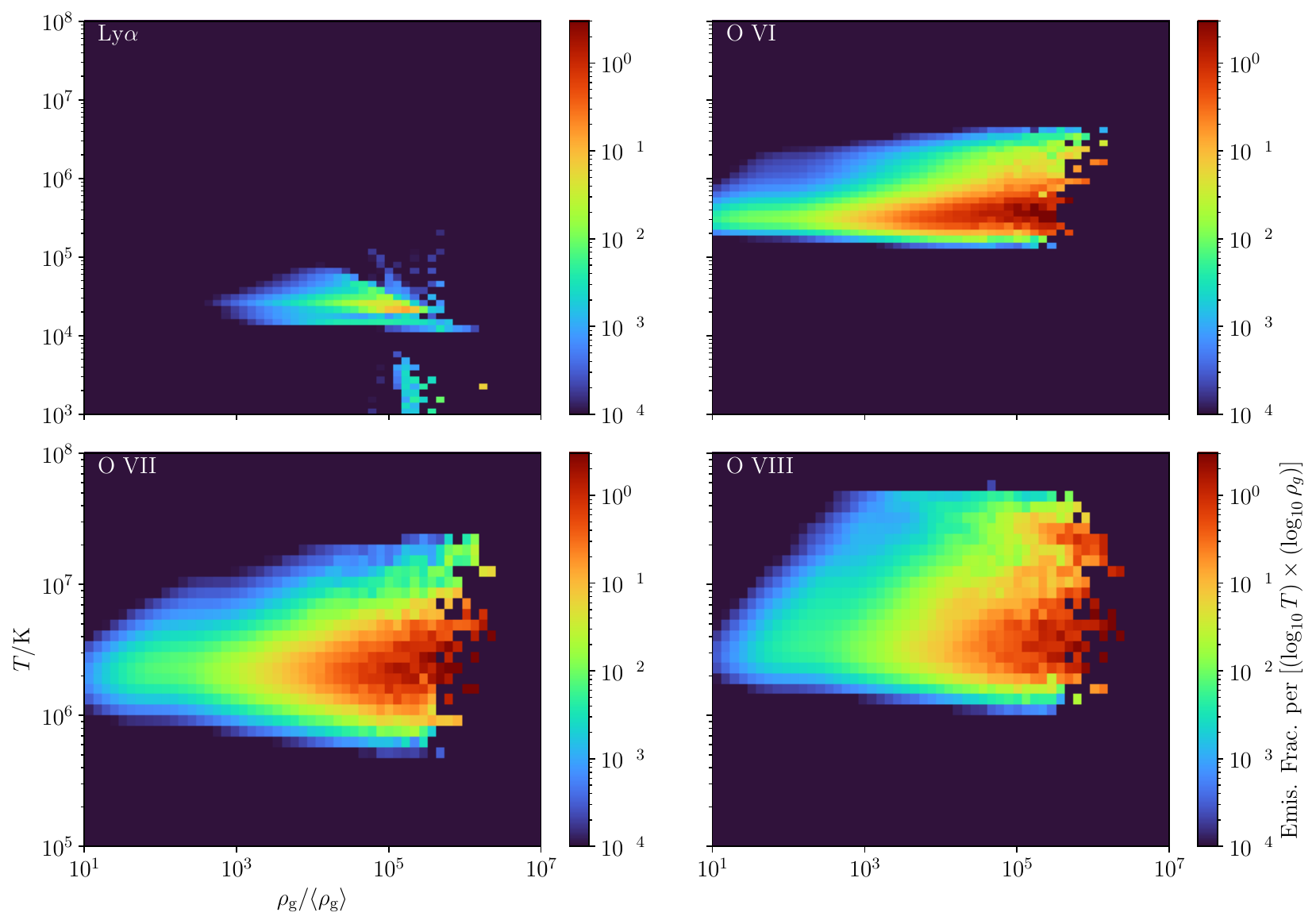}
  \caption{Distribution of emissivities of different
    emission lines in the
    $\{\log_{10}T\times \log_{10}\rho_{\rm g}\}$ phase
    space, indicating the relative fractions of emission
    power per dex squared. The orange, green, and blue
    curves in each panel indicate the contours for the low
    resolution run, at which the distribution functions take
    the values $10^{-2}$, $10^{-1}$, and $10^0$,
    respectively. } 
  \label{fig:hist_Trho} 
\end{figure*}

\begin{figure*}
  \centering
  \hspace*{-0.5cm}
  \includegraphics[width=7.2in, keepaspectratio]
  {\figdir/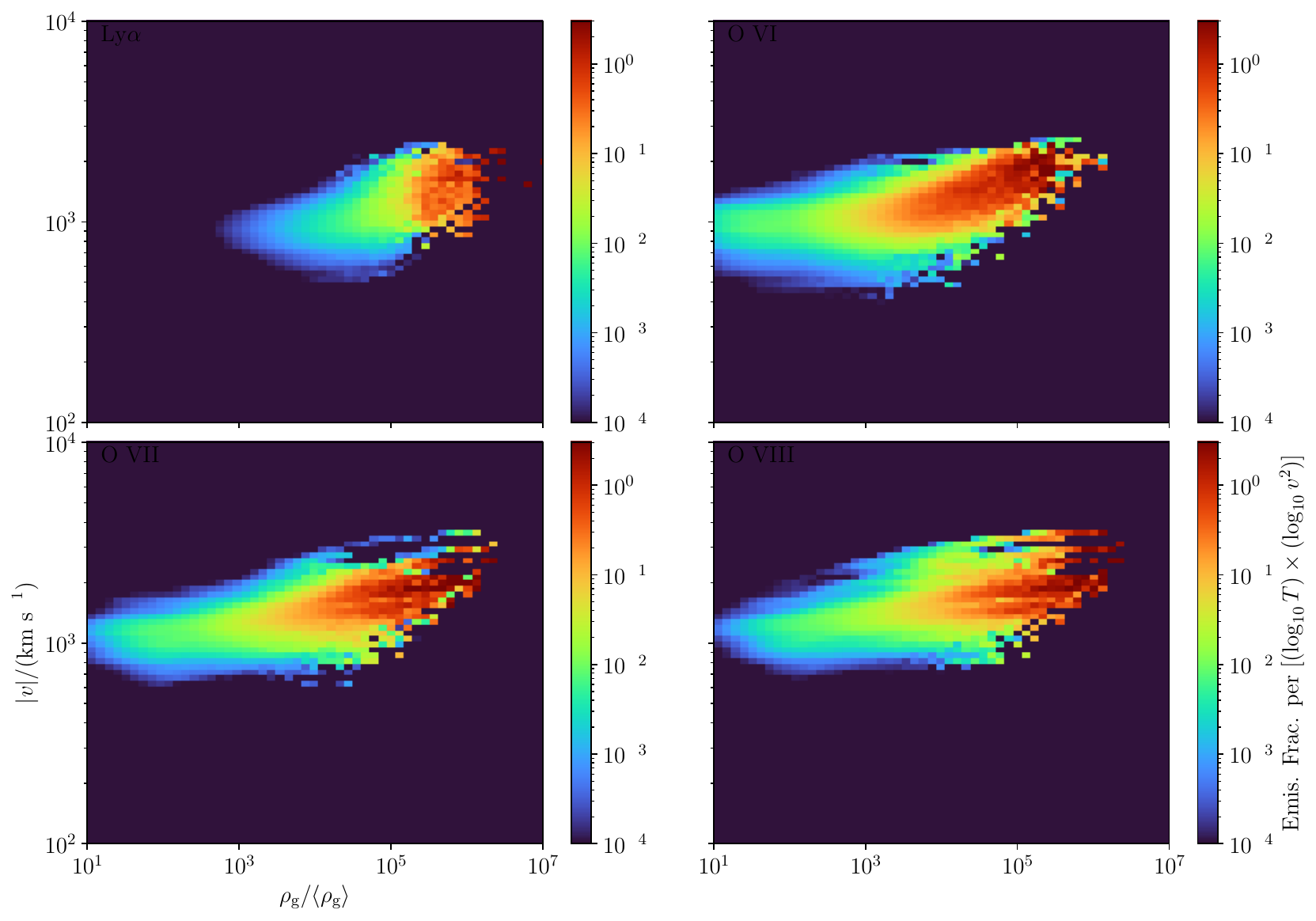}
  \caption{Similar to Figure~\ref{fig:hist_Trho}, but shows
    the distribution functions in the
    $\{\log_{10}|v|\times \log_{10}\rho_{\rm g}\}$ phase
    space. }
  \label{fig:hist_v2rho} 
\end{figure*}
 
A principal objective of the grid-based high-resolution
cosmological simulations using \kratos{} is to generate
synthetic mock observations that inform observational
strategies for detecting the WHIM gas. This diffuse plasma,
comprising $\sim 23.4\%$ of the universe's baryons at $z=0$,
remains challenging to observe due to its low density
($n \sim 10^{-6}~\mathrm{cm}^{-3}$) and high ionization
state. To address this, our mock observations focus on four
key emission tracers:
\begin{itemize}
\item \lya{}: Hydrogen line at $1216$~\AA in FUV, sensitive to
 gas near $T \sim 10^4~\K$;
\item \ion{O}{vi} $\lambda\lambda 1032, 1038$~\AA: Resonance
  doublet probing collisionally ionized oxygen
  in $T \sim 10^{5.5}~\K$ gas, extensively observed in
  circumgalactic gas \citep{2011Sci...334..948T};
\item \ion{O}{vii} $\lambda 21.6$~\AA: The He-like
  triplet ($\chem{O^{6+}}$) from collisional excitation and
  recombination in $T \sim 10^{6.5}~\K$ plasma, a predicted
  signature of the X-ray WHIM
  \citep{2002ApJ...564..604F, 2009ApJ...697..828B};
\item \ion{O}{viii} $\lambda 19.0$~\AA: X-ray line
  ($\mathrm{O^{7+}}$) dominant in $T > 10^{7}~\K$ gas.
\end{itemize}

As illustrated in Figure~\ref{fig:cosmic_render}, these
tracers should exhibit different morphological
correlations. Ly$\alpha$ and \ion{O}{vi} emission
preferentially trace the $\sim 0.1-1~\Mpc$ wide ``cool
spines'' of cosmic filaments ($T \lesssim 10^5~\K$), where
gas densities ($n \sim 10^{-4}~\cm^{-3}$) sustain
collisional ionization equilibrium for \ion{O}{vi}.
Conversely, \ion{O}{vii} and \ion{O}{viii} X-ray emission
predominantly originate from the virialized gas in and
around halos with $M \gtrsim 10^{12}~M_\odot$, where the
virial temperature (eq.~\ref{eq:T_vir}) exceeds
$10^6~\K$. These halos host accretion shocks that thermalize
infalling gas to X-ray temperatures.

\begin{figure}
  \centering
  \includegraphics[width=3.25in, keepaspectratio]
  {\figdir/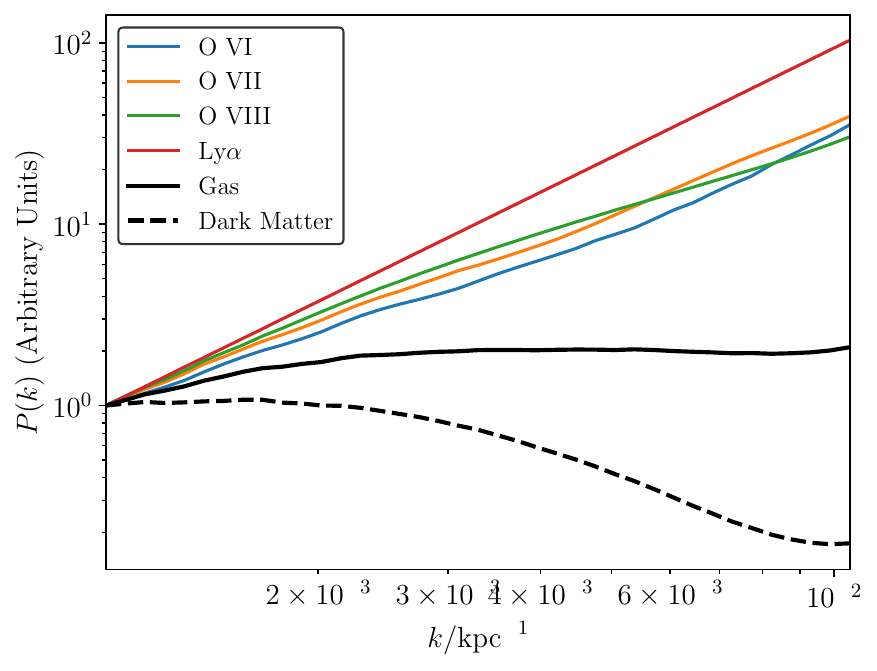}
  \caption{The two-dimensional power spectra of surface
    brightness of \lya{}, \ion{O}{vi}, \ion{O}{vii}, and
    \ion{O}{viii} emission lines, and the projected
    densities (gas and dark matter) over the $100~\Mpc~h^{-1}$
    depth.} 
  \label{fig:powerspec_2d} 
\end{figure}

\subsection{\lya{} and Cosmic Webs} 
\label{sec:result-obv-lya}

The Lyman $\alpha$ (\lya{} hereafter) emissivity based on the
fiducial simulation is calculated based on two processes,
including the collisional excitation of neutral hydrogen
(see e.g., \citealt{DraineBook}), and the recombination to
the state with principal quantum number $n=2$ using the
fitting formula in \citet{2019PhyS...94e5403K} covering the
temperature up to $T > 10^7~\K$. As demonstrated in multiple
studies \citep[e.g.][]{2001ApJ...552..473D,
  2006ApJ...650..560C} and visualized in
Figure~\ref{fig:cosmic_render}, \lya{} emission serves as a
critical diagnostic for probing the warm-hot intergalactic
medium (WHIM) within cosmic filaments. This capability
arises from Ly$\alpha$'s sensitivity to both recombination
processes in ionized hydrogen and resonant scattering in
partially ionized gas. The observed filamentary morphology
aligns with predictions from cosmological hydrodynamical
simulations which identify two distinct density regimes
traced by Ly$\alpha$. Throughout this paper $\rho_{\rm
  c1}$ denotes the critical density evaluated with
$H_0=100~\km~\s^{-1}~\Mpc^{-1}$ (i.e.\ $h=1$),
$\rho_{\rm c1}\approx1.1\times10^{-5}~m_p~\cm^{-3}$. One
focuses on the high-density inflows
($\rho \sim 10^1- 10^3~\rho_{\rm c1} \sim 10^{-4}-
10^{-2}~m_p~\cm^{-3}$), which follow the gas accreted onto
massive halos ($M \gtrsim 10^{12}~M_\odot$) through cold
streams. The other is the diffuse cosmic web
($\rho \sim \rho_{\rm crit} \approx 10^{-6}~m_p~\cm^{-3}$),
pervasive filamentary networks connecting virialized
structures. Such bimodality can also be identified in
Figures~\ref{fig:hist_Trho} and \ref{fig:hist_v2rho}.  The
higher-resolution grid-based simulations (spatial resolution
$\Delta x \sim 24.5~\kpc~h^{-1}$) resolve and reveal the clumpy
accretion shocks with Mach numbers
$\mathcal{M} \sim 2 - 10$ (weaker than the
$\mathcal{M} \sim 10-100$ accretion shocks onto massive
halos; \S\ref{sec:results-thermal}), and turbulent
interfaces that are led by Kelvin-Helmholtz instabilities at
filament-void boundaries.

The critical role of resolution is quantified in
Figures~\ref{fig:powerspec_2d}. While low-resolution
simulations ($\Delta x > 10^2~\mathrm{kpc}$) fail to resolve
densities $\rho_{\rm g} \gtrsim 10^2\rho_{\rm c1}$,
our fiducial resolution keeps the non-linear power
spectrum $P(k)$ undamped up to
$k \sim 0.1~\kpc^{-1}$ (spatial scales $\sim 20~\kpc$). This
resolution dependence arises from the Jeans criterion,
\begin{equation}
  \begin{split}
  \lambda_J & \equiv \left(\dfrac{\pi c_s^2}{G \rho}
  \right)^{1/2} \\
    & \simeq 20~\kpc\times 
  \left(\dfrac{T}{10^4~\K}\right)^{1/2}
  \left(\dfrac{\rho}{10^{-2}m_p~\cm^{-3}}\right)^{-1/2},
  \end{split}
\end{equation}
requiring $\Delta x \lesssim \lambda_J$ to assure that the
collapse of self-gravitating clumps can take place, which is
essential for modeling baryon cycling between filaments and
halos.

\begin{figure*}
  \centering
  \hspace*{-0.7cm}
  \hspace{0.03\textwidth}
  \includegraphics[width=0.4\textwidth, keepaspectratio]
  {\figdir/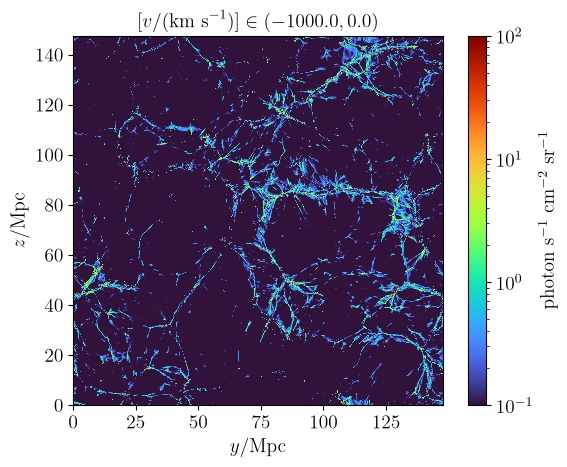}
  \includegraphics[width=0.4\textwidth, keepaspectratio]
  {\figdir/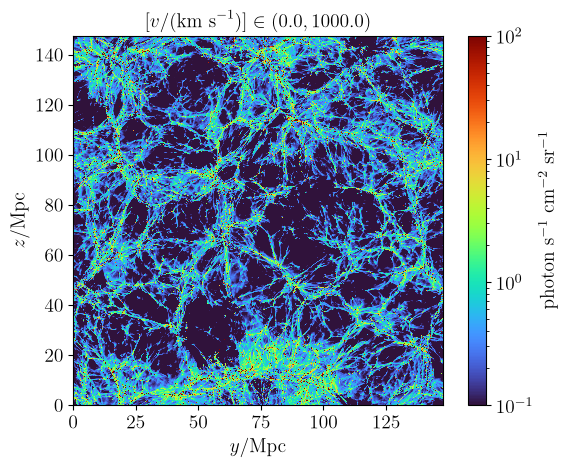} \\
  \includegraphics[width=0.4\textwidth, keepaspectratio]
  {\figdir/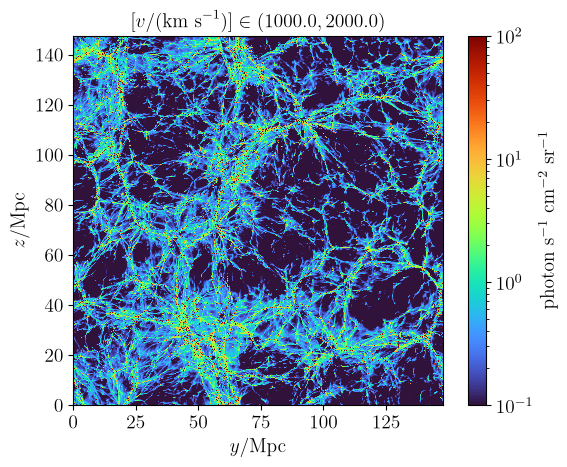}  
  \includegraphics[width=0.4\textwidth, keepaspectratio]
  {\figdir/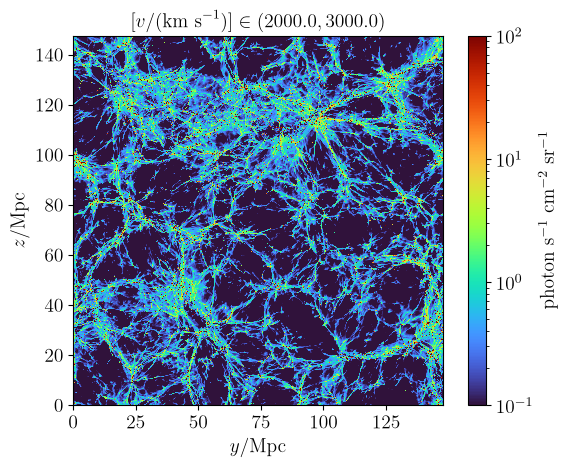} \\
  \includegraphics[width=0.4\textwidth, keepaspectratio]
  {\figdir/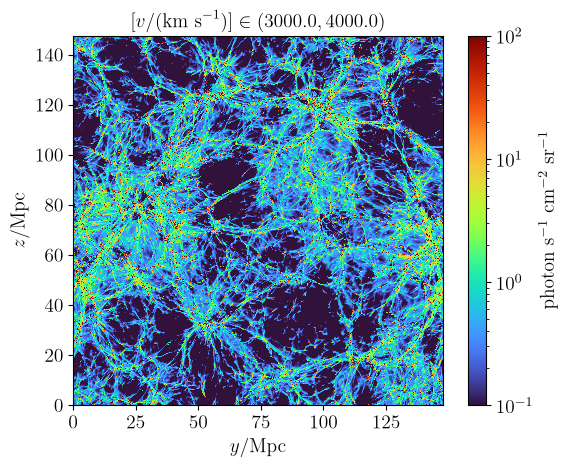}
  \includegraphics[width=0.4\textwidth, keepaspectratio]
  {\figdir/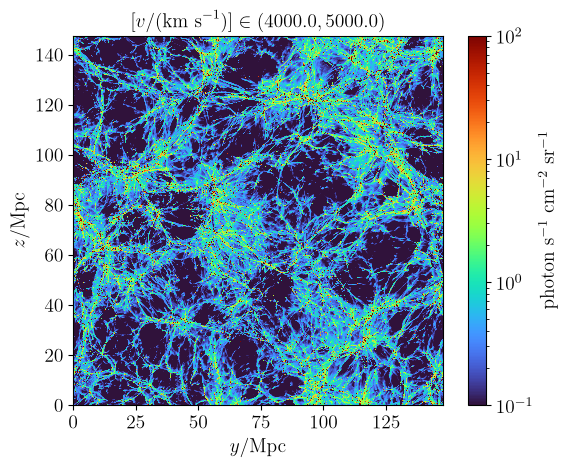} \\
  \includegraphics[width=0.4\textwidth, keepaspectratio]
  {\figdir/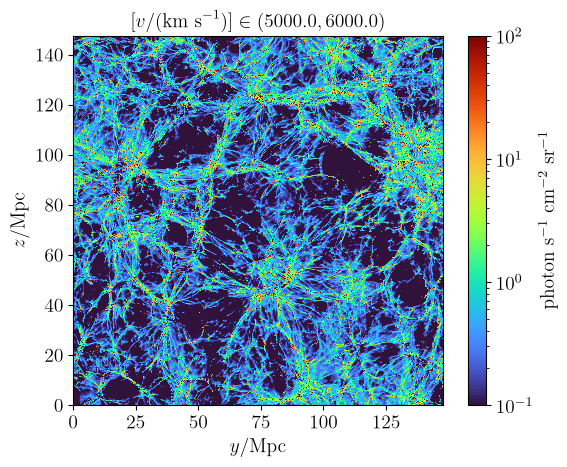}
  \includegraphics[width=0.4\textwidth, keepaspectratio]
  {\figdir/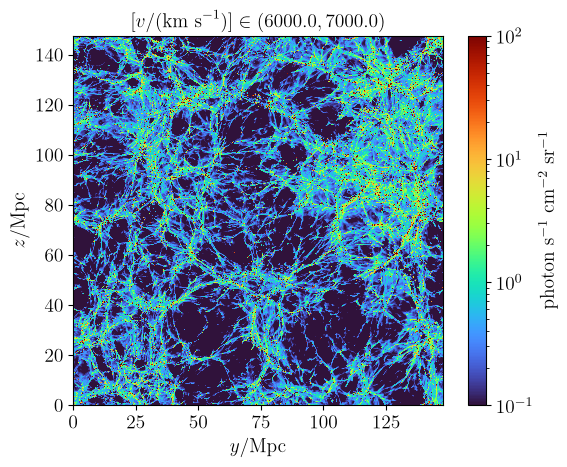}
  \caption{Similar to Figure~\ref{fig:emis_sb}, but shows
    the surface brightness in each bin in the line-of-sight
    velocity space (indicated at the top of each
    panel). Note that the cosmological redshifts based on
    the distance to the observer is taken into account. }
  \label{fig:lya_channel} 
\end{figure*}

The velocity-resolved channel maps presented in
Figure~\ref{fig:lya_channel} provide insights into the
kinematic structure of 
WHIM. 
%By accounting for cosmological redshift effectsthrough velocity slicing (
With a thickness of $\Delta v = 500~\km~\s^{-1}$ per channel, these maps resolve the full 3D structures of
Ly$\alpha$ emission, revealing the coherent motions
($v \sim 200-500~\km~\s^{-1}$) along filamentary axes
(driven by gravitational infall into cosmic web nodes), and
broader turbulent motions.

Comparison with the integrated surface brightness map
(Figure~\ref{fig:emis_sb}) demonstrates that velocity
slicing enhances filament detectability significantly. This
technique isolates structures through their Doppler and
cosmological shifts, mitigating confusion from foreground
and background emission. The recovered filaments
(Figure~\ref{fig:lya_channel}) exhibit characteristic
physical parameters,
$\rho \gtrsim 10^{-2}\rho_{\rm c1} \approx
10^{-7}~m_p~\cm^{-3}$ {\it (i.e.\ the Ly$\alpha$-bright
filaments trace gas at and above the cosmic mean baryon
density---the Ly$\alpha$ forest---rising above the much more
diffuse, photoionized intergalactic background)}, and
$T < 10^4~\K$, being consistent
with Ly$\alpha$ excitation and efficient recombination to
the $n = 2$ states. These measurements align with
predictions from structure formation simulations
\citep{1999ApJ...514....1C, 2001ApJ...552..473D,
  2006ApJ...650..560C}, where filaments represent shocked
accretion flows with Mach numbers $\mathcal{M} \sim 2-10$.

The resolved kinematics have significant implications for
future UV missions focusing on emissions, which require
spectral resolution
$R \equiv \lambda/\Delta\lambda \gtrsim 1000$
($\Delta v < 300~\km~\s^{-1}$) to disentangle filament
structures from the integrated field. In addition,
cross-correlation with X-ray and optical observations with
high-spatial-resolution ($\lesssim 1~{\rm arcsec}$) surveys
needed to match filament angular scales. Upcoming facilities
like the \textit{Habitable Worlds Observatory} (UV) and
Athena (X-ray) will test these predictions through
tomographic mapping of Ly$\alpha$ forest transmission
fluctuations, and velocity-resolved X-ray emission line
spectroscopy of \ion{O}{vii} and \ion{O}{viii} elaborated in
what follows.

\subsection{Emission from Oxygen Ions}
\label{sec:result-obv-oxy}

According to the observational diagnostics presented in
Figures~\ref{fig:emis_sb}, \ref{fig:hist_Trho}, and
\ref{fig:hist_v2rho}, the intergalactic gas traced by oxygen
ion emission lines exhibits distinct thermodynamic and
kinematic properties compared to gas traced by \lya{}
emission. The oxygen abundance adopted for these emission
calculations follows the imposed metallicity field of
\S\ref{sec:method-metals}. The X-ray emission features of \ion{O}{vii} and
\ion{O}{viii} predominantly trace gas in the vicinity of
halos. In the velocity--density phase space
(Figure~\ref{fig:hist_v2rho}) the mass-weighted speed $|v|$
of this emitting gas peaks at $|v|\sim 250~\km~\s^{-1}$.
This peak coincides with the virial velocity of the host
halos: a halo of mass $M\sim 10^{12}$--$10^{13}~M_\odot$ has
$V_{\rm vir}\simeq 140$--$310~\km~\s^{-1}$, so the
intergalactic gas that falls in, collides with the halo
virialized medium, and shock-heats to $T\gtrsim 10^6~\K$
moves at comparable velocities. These shocks are mostly localized
around halos rather than large-scale cosmic web filaments.
%as the latter exhibit shallower potential wells and lower virial velocities. 
Furthermore, the high-density post-shock
gas ($\rho \sim 10^2-10^3~\rho_{\rm c1}$) occupies the
high-velocity tail of the distribution at
$|v|\sim 450~\km~\s^{-1}$ (Figure~\ref{fig:hist_v2rho}): this
three-dimensional speed is reached toward the deeper
potential wells of the most massive
($M\sim 10^{13}$--$10^{14}~M_\odot$) halos, where the radial
infall and the internal velocity dispersion combine. Such
elevated speeds and densities require high-resolution hydrodynamical
simulations to resolve, as they involve sub-kpc scale
processes such as shock front instabilities and turbulent
mixing.
 
The behavior of the ultraviolet \ion{O}{vi}
$\lambda\lambda$1032, 1038 emission lines presents a more
complex picture compared to the other oxygen tracers. In the
$\{\log_{10}|v|\times \log_{10}\rho_{\rm g}\}$ phase space
(Figure~\ref{fig:hist_v2rho}), the emissivity distribution
of \ion{O}{vi} still overlaps largely with those of
\ion{O}{vii} and \ion{O}{viii}---indicating similar
bulk-flow kinematics---but is centered at somewhat lower
densities, because \ion{O}{vii} and \ion{O}{viii} originate
in the densest, hottest post-shock gas deep inside halos,
whereas \ion{O}{vi} emits efficiently from the more extended
gas surrounding them. In the
$\{\log_{10}T\times \log_{10}\rho_{\rm g}\}$ phase space
(Figure~\ref{fig:hist_Trho}) and in configuration space the
segregation is clearer: the \ion{O}{vi}-emitting gas occupies
a cooler locus ($T\sim 10^{5.5}-10^6~\K$), and---since the
emissivity scales as $\rho_{\rm g}^2$---its emission is
dominated by the densest structures available at those
temperatures, i.e.\ it picks out overdense interface regions
relative to the diffuse photoionized background. This
thermodynamic segregation is expected, because the lower
ionization state preferentially occupies cooler gas, while
\ion{O}{vii} and \ion{O}{viii} require the higher
temperatures characteristic of virialized halo gas
($T > 10^6~\K$). In the meantime,
the \ion{O}{vi} $\lambda\lambda$1032, 1038 doublet is a
Li-like transition with an excitation energy of only
$\sim 12$~eV, versus $\sim 0.6$~keV for the K-shell X-ray
lines of \ion{O}{vii} and \ion{O}{viii}; the far gentler
collisions that excite \ion{O}{vi} therefore allow its
emission to be produced throughout the cooler, more extended
halo outskirts, rather than solely in the hot post-shock
cores. This
extended emission traces the interface zones where
shock-heated WHIM gas at $T\sim 10^{5.5}-10^6$ K mixes with
cooler filamentary gas, potentially mapping the multiphase
structure of the cosmic web on the transitional regions
between the warm Ly$\alpha$ forest and the hot X-ray
emitting intracluster medium.

\begin{figure*}
  \centering
  \hspace*{-0.7cm}
  \hspace{0.03\textwidth}
  \includegraphics[width=0.4\textwidth, keepaspectratio]
  {\figdir/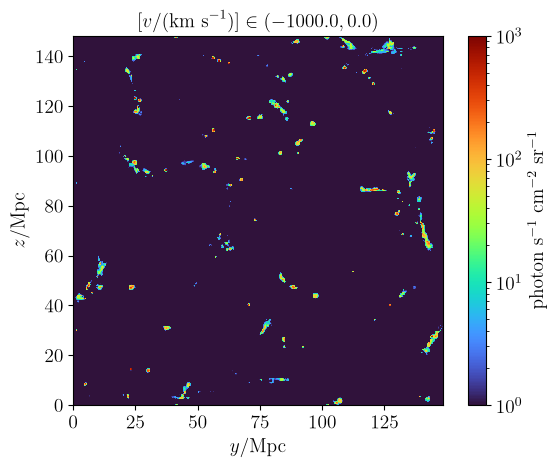}
  \includegraphics[width=0.4\textwidth, keepaspectratio]
  {\figdir/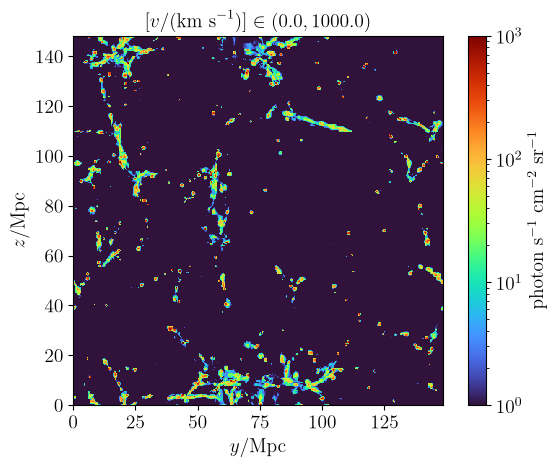}  \\
  \includegraphics[width=0.4\textwidth, keepaspectratio]
  {\figdir/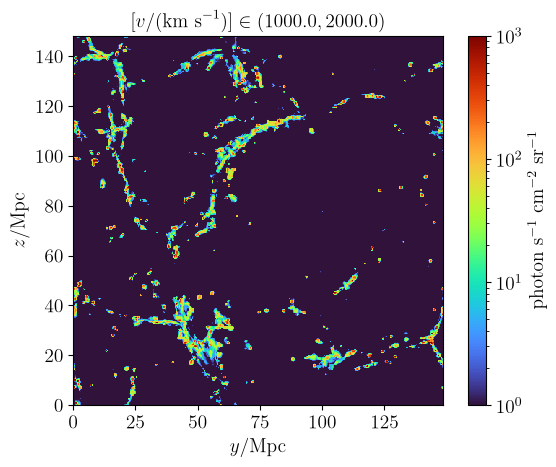}  
  \includegraphics[width=0.4\textwidth, keepaspectratio]
  {\figdir/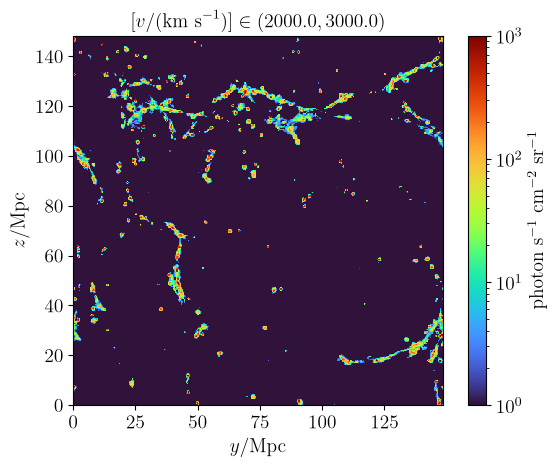}  \\
  \includegraphics[width=0.4\textwidth, keepaspectratio]
  {\figdir/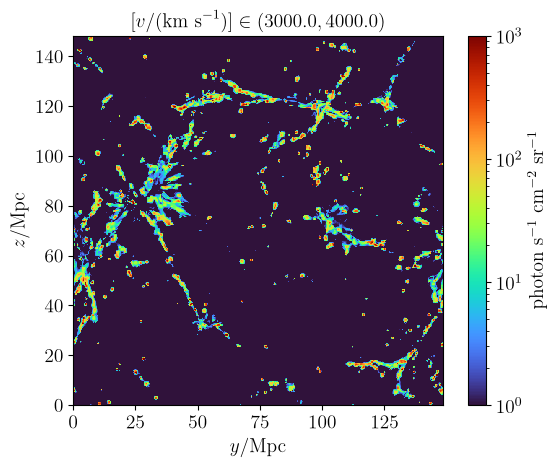}
  \includegraphics[width=0.4\textwidth, keepaspectratio]
  {\figdir/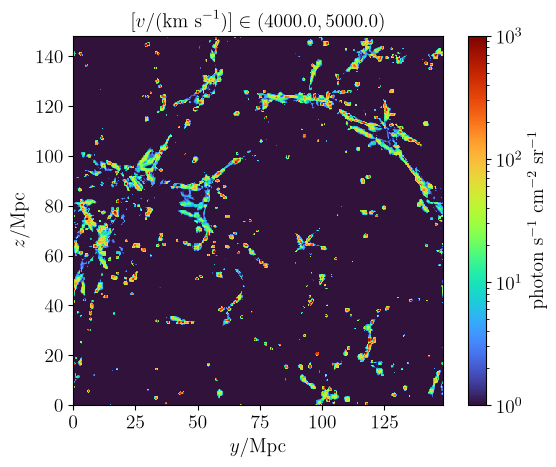}  \\
  \includegraphics[width=0.4\textwidth, keepaspectratio]
  {\figdir/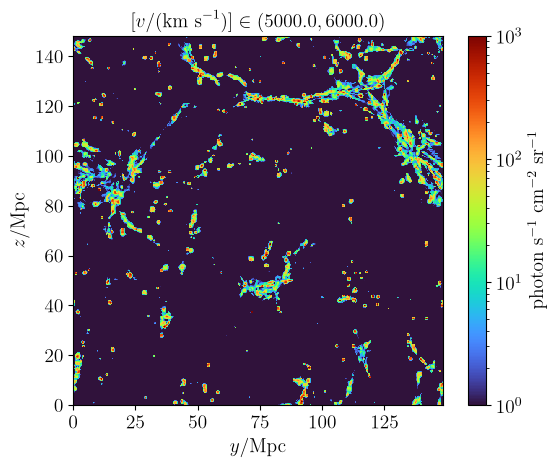}
  \includegraphics[width=0.4\textwidth, keepaspectratio]
  {\figdir/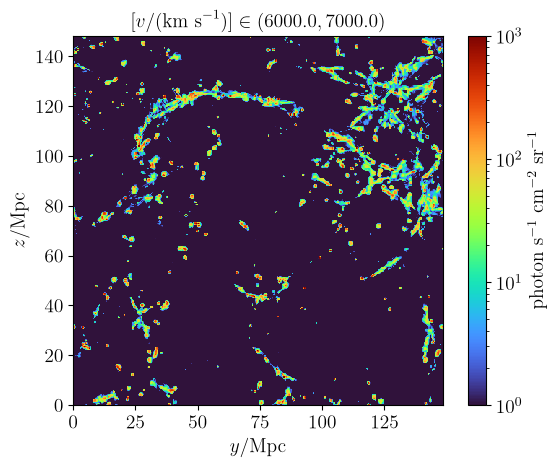}
  \caption{Similar to Figure~\ref{fig:lya_channel}, but
    plotted for the \ion{O}{vi} $\lambda\lambda$1032, 1038
    emissions (note also that the upper limit of colormaps
    are also different). }
  \label{fig:o6_channel} 
\end{figure*}

The zoom-in Figure~\ref{fig:emis_o6_zoomin} panel
corroborates this by showing that the diffuse \ion{O}{vi}
emission traces the interface regions between massive halos
($M > 10^{13}~M_\odot$) and the densest filaments of the
cosmic web, where shock-heated gas mixes with cooler,
inflowing material. When disentangled by different velocity
channels (Figure~\ref{fig:o6_channel}), the densest spines
of the cosmic web are more prominently illustrated, which is
helpful in identifying the overall multidimensional
structures and kinematic behaviors in the cosmic webs in
conjunction with other methods such as \lya{}. This
velocity-resolved analysis reveals coherent infall motions
along filament axes ($v\sim 200-500$ km s$^{-1}$) and
broader turbulent motions, providing crucial insights into
the accretion processes that build the cosmic web.

Future multi-wavelength observational campaigns will be
helpful for mapping the multiphase structure of the
warm-hot intergalactic medium (WHIM). Combining X-ray
(\ion{O}{vii}, \ion{O}{viii}) and ultraviolet (\ion{O}{vi},
\lya{}) tracers will enable reconstruction of the gas
density, temperature, and velocity fields across different
spatial scales. For instance, X-ray microcalorimeters aboard
missions like ATHENA and HUBS reach an instrumental velocity
resolution of $\Delta v \sim 10^3~\km~\s^{-1}$ at the
\ion{O}{vii} line ($E\simeq 0.574$~keV), while XRISM reaches
$\Delta v \sim 2-3\times 10^3~\km~\s^{-1}$. These figures are
large in velocity units only because of the low
($\lesssim 1$~keV) line energy, not because of any intrinsic
gas motion. Since the characteristic WHIM inflow and
turbulent speeds ($|v|\sim 250$--$450~\km~\s^{-1}$;
\S\ref{sec:result-obv-oxy}), as well as the thermal width of
the lines ($\lesssim 200~\km~\s^{-1}$ at $T\lesssim
10^7~\K$), lie below these resolutions, the \ion{O}{vii}
profiles of WHIM shocks will be only marginally resolved:
these instruments will chiefly constrain line centroids,
equivalent widths, and the highest-velocity infall wings,
rather than the detailed velocity substructure.
Concurrently, UV spectrographs such as the
\textit{Habitable Worlds Observatory} could also be applied
to detect \ion{O}{vi} kinematics in filament-halo
interfaces, observationally constraining gas outflows and
halo accretion rates. Cross-correlating these datasets with weak
gravitational lensing surveys will further disentangle the
contributions of baryonic physics and dark matter potential
wells to the observed emission. Such synergies are essential
for advancing our understanding of the cosmic baryon cycle
and the role of the WHIM in galaxy evolution.

\section{Discussion and Summary}
\label{sec:summary}

We have presented a suite of high-resolution cosmological
simulations using the novel GPU-accelerated code \kratos{}
to investigate the formation, evolution, and observable
signatures of the Warm-Hot Intergalactic Medium (WHIM)
gas. By resolving gas dynamics down to
$\Delta x\sim 24.5~\kpc~h^{-1}$ scales in a $(100~\Mpc~h^{-1})^3$ comoving
volume, our simulations self-consistently capture the
thermalization of baryons through accretion shocks, the
multiphase structure of the cosmic web, and the long-term
survival of WHIM against radiative cooling. A simplified
density- and redshift-dependent metal-enrichment
prescription (\S\ref{sec:method-metals}) enables the
synthetic oxygen-line emission maps of
\S\ref{sec:results-obv}. 
Neglecting AGN or star formation feedback heating, around
$\sim 23.4\%$ of all baryons reside in the WHIM phase
($10^5<T/\K < 10^7$) by $z=0$, predominantly located in the
filaments and around massive
($M \gtrsim 10^{13}~M_\odot$) dark matter halos. 
This fraction is substantially lower than the $40-50\%$ found in earlier, lower-resolution simulations. Our resolution study (Figure~\ref{fig:frac-whim}; Appendix~\ref{sec:resolution-analyses}) demonstrates that spatial resolution plays a pivotal role in determining the WHIM fraction: resolving gas near its Jeans scale allows it to reach higher densities, where enhanced radiative cooling transfers a substantial fraction of baryons out of the WHIM temperature range.
A resolution of $50~h^{-1}~{\rm kpc}$ is required to correctly capture the WHIM fraction, as our study shows, compared to the $100$--$200~h^{-1}~{\rm kpc}$ resolution employed in \citet{1999ApJ...514....1C}.

%This gas accumulates through hierarchical structure formation, withits thermodynamic history shaped by multiple episodes of shock heating during mergers and cosmic web assembly. 
%The characteristic density ($\rho\sim 10^0-10^2~\rho_{\rm c1}$) and temperature ($10^{5.5}\lesssim T/\K \lesssim 10^{6.5}$) ranges found in our simulations align with theoretical expectations for the ``missing baryon'' reservoir, while the resolved Bondi radii around $M\gtrsim 10^{11}~M_\odot$ halos enable detailed modeling of gas accretion processes.

The synthetic observations reveal distinct morphological and
kinematic signatures across tracers: \lya{} and \ion{O}{vi} emission preferentially trace the
$\sim 0.1~\Mpc$ spine filaments and halo interfaces where
turbulence and shocks ($\Delta v \gtrsim 200~\km~\s^{-1}$)
mix warm and hot phases, while \ion{O}{vii} and
\ion{O}{viii} X-ray lines predominantly originate from
virialized gas in massive halos. The velocity-resolved
channel maps demonstrate that future observatories with
spectral resolution $R> 10^3$ will be essential for
disentangling the kinematic substructure of WHIM
filaments. These predictions provide critical guidance for
upcoming missions like XRISM, ATHENA, and HUBS, whose
microcalorimeters will probe the thermal and chemical state
of WHIM through \ion{O}{vii} and \ion{O}{viii} line
diagnostics, while UV spectrographs like the \textit{Habitable Worlds Observatory} could map
the \ion{O}{vi}-bright interfaces between filaments and
halos.

Future work will focus on forward-modeling these simulations
into mock survey data products that incorporate instrumental
response functions, foreground contamination, and cosmic
variance. Direct comparisons with XRISM and HUBS
commissioning data will test our predictions for WHIM
density and temperature distributions and oxygen
abundances. 
Extending the simulations to include
magnetohydrodynamics %and anisotropic thermal conduction
will
address current limitations in modeling the WHIM's observables, such as synchrotron emission.
%magnetization and plasma instabilities. 
%Additional parameterstudies exploring AGN feedback prescriptions and early ($z\gtrsim 3$) heating sources are needed to resolvelingering tensions between the simulated \lya{} forest opacity and observations. 
%By coupling $\Delta x = 24.5~\kpc~h^{-1}$ resolution simulations with multi-wavelength observational synthetics, this framework will enable precise tests of theWHIM's role in the cosmic baryon cycle and its connection to galaxy formation physics.

\begin{acknowledgments}
We acknowledge the support from the start-up funding of Zhejiang University and Zhejiang provincial top level research support program. The simulations and analysis presented in this article were carried out on the SilkRiver Supercomputer of Zhejiang University and the ``Zimo'' high-performance computing cluster of the Purple Mountain Observatory of the Chinese Academy of Sciences. The authors used the DeepSeek large language model \citep{deepseek2025} to polish the language and improve the clarity of the manuscript text; all scientific content, analysis, and conclusions remain the responsibility of the authors.
\end{acknowledgments}

\bibliographystyle{aasjournal}
\bibliography{whim}

\appendix 

\section{Resolution Analyses}
\label{sec:resolution-analyses}

\subsection{The WHIM Mass Fraction and Its Phase-Space
  Origin}

To quantify the impact of numerical resolution on the WHIM
census, we examine the gas mass distribution function over
temperature for the full resolution series and for the
rebinned $N=4096$ output (Figure~\ref{fig:frac-whim},
presented in \S\ref{sec:results-thermal}).  The
full-resolution and rebinned $N=4096$ curves agree at the
extremes ($T\lesssim 10^{4}~\K$ and $T\gtrsim 10^{7}~\K$)
where the extensive quantities (mass, momentum, energy, and
entropy densities) are conserved under rebinning, but
diverge in the intermediate WHIM regime
($10^{4}\lesssim T/\K \lesssim 10^{7}$) because temperature,
as an intensive quantity, is a non-linear combination of the
conserved fields. The WHIM mass fraction decreases from
$51.4\%$ at $N=512$ and $36.4\%$ at $N=1024$ to $21.6\%$ at
$N=2048$, and is $23.4\%$ for the fiducial $N=4096$ run,
while rebinning the $N=4096$ output by a factor of four
inflates it back to $46.7\%$, close to the $N=512$ value.
The slight non-monotonic rise from $21.6\%$ at $N=2048$ to
$23.4\%$ at $N=4096$ ($+1.8$~pp) can be attributed to two
physical effects rather than a sign of non-convergence:
(i)~large hydrodynamic simulations are subject to
substantial non-linearity that manifests as stochasticity
between resolution levels, and (ii)~higher resolution better
resolves shock-heating processes, legitimately enhancing the
production of WHIM-temperature gas. The overall trend---a
decline from $51.4\%$ at $N=512$ to $\sim 22\!-\!23\%$ at
$N\geq 2048$ followed by near-stabilization---indicates that
the WHIM fraction is approaching convergence at the
$\sim 2$~pp level.  The low-resolution runs deviate from the
fiducial result at all $T\gtrsim 10^{4}~\K$, as they fail to
capture gravitational collapse and subsequent catastropic,
almost irreversible cooling driven by Jeans instability,
leaving too much gas at WHIM temperatures that should have
collapsed into denser phases or cooled below $10^5~\K$. Full
spatial resolution is essential for accurately modeling the
thermodynamic evolution of the intergalactic medium.  The
divergence between the full and rebinned $N=4096$ curves in
the WHIM temperature range demonstrates that the higher
resolution captures critical physical structures, filaments,
shocks, and cooling interfaces, that are absent from the
coarser effective grid, even when the coarser grid inherits
strictly conserved density and energy fields.

% We note that the emergence of cool ($\sim 10^4~\K$)
% compact regions at $N\geq 1024$ could, in principle,
% partly reflect numerical overcooling---a known failure
% mode of grid-based codes when the cooling length is
% marginally resolved.  A cooling-off control run would be
% needed to unambiguously distinguish genuine gravitational
% collapse from this artifact; such a test is left for
% future work.

\begin{figure*}
  \centering
  \includegraphics[width=0.92\textwidth, keepaspectratio]
  {\figdir/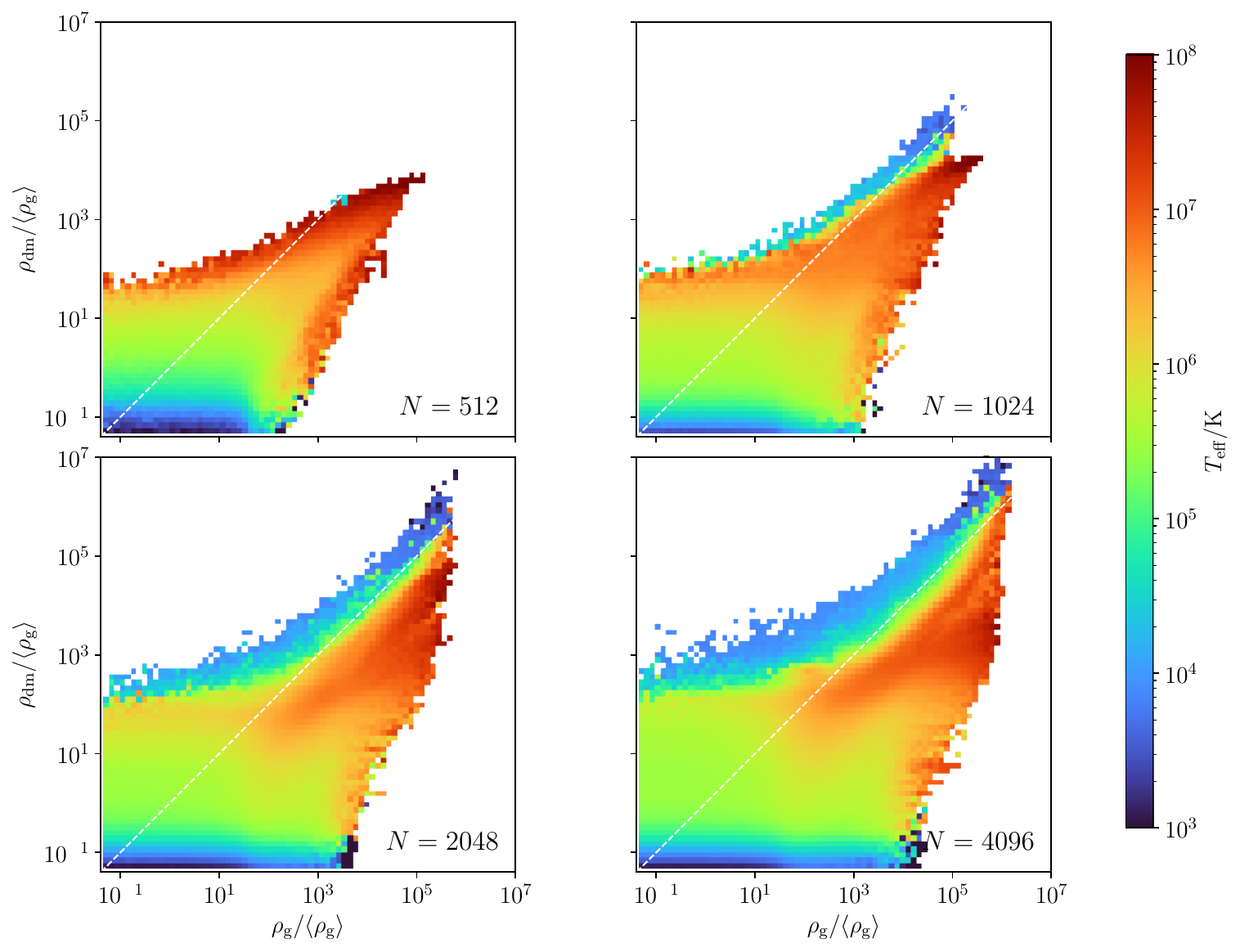}
  \caption{Mass-weighted effective temperature
    $T_{\rm eff}\equiv(\langle p\rangle/\langle\rho_{\rm
    g}\rangle)\,m_p/(2k_{\rm B})$ (the ionized limit) in the
    gas--dark-matter density plane at $z\simeq 0$ for
    $N=512$ (upper left), 1024 (upper right), 2048 (lower
    left), and 4096 (lower right). Both densities are
    normalized to the mean gas density, and the white dashed
    line marks $\rho_{\rm g}=\rho_{\rm dm}$. With increasing
    resolution the populated phase space extends to higher
    overdensities, the shocked plume along the
    $\rho_{\rm g}\sim\rho_{\rm dm}$ diagonal
    ($T_{\rm eff}\sim 10^6$--$10^8~\K$) becomes more
    pronounced, and cold ($T_{\rm eff}\sim 10^3$--$10^4~\K$)
    dense gas at the highest dark-matter overdensities is
    resolved only for $N\geq 2048$. }
  \label{fig:phase-res}
\end{figure*}

The phase-space origin of this trend is illustrated in
Figure~\ref{fig:phase-res}, which maps the mass-weighted
effective temperature in the plane of gas versus dark-matter
density. With increasing resolution the simulation populates
increasingly overdense gas, the shocked plume along
$\rho_{\rm g}\sim\rho_{\rm dm}$ at $T_{\rm eff}\sim
10^{6}$--$10^{8}~\K$ becomes more pronounced, and cold
($10^{3}$--$10^{4}~\K$) dense gas associated with the
strongest dark-matter concentrations emerges only for
$N\geq 2048$. The excess WHIM in the under-resolved runs
stems precisely from gas that is artificially held at
$10^{5}$--$10^{7}~\K$ because these cold, dense phases and
the thin cooling interfaces around them are not resolved.

\subsection{Phase-Space Cooling Signatures and the Jeans
  Scale}

\begin{figure*}
  \centering
  \includegraphics[width=0.94\textwidth, keepaspectratio]
  {\figdir/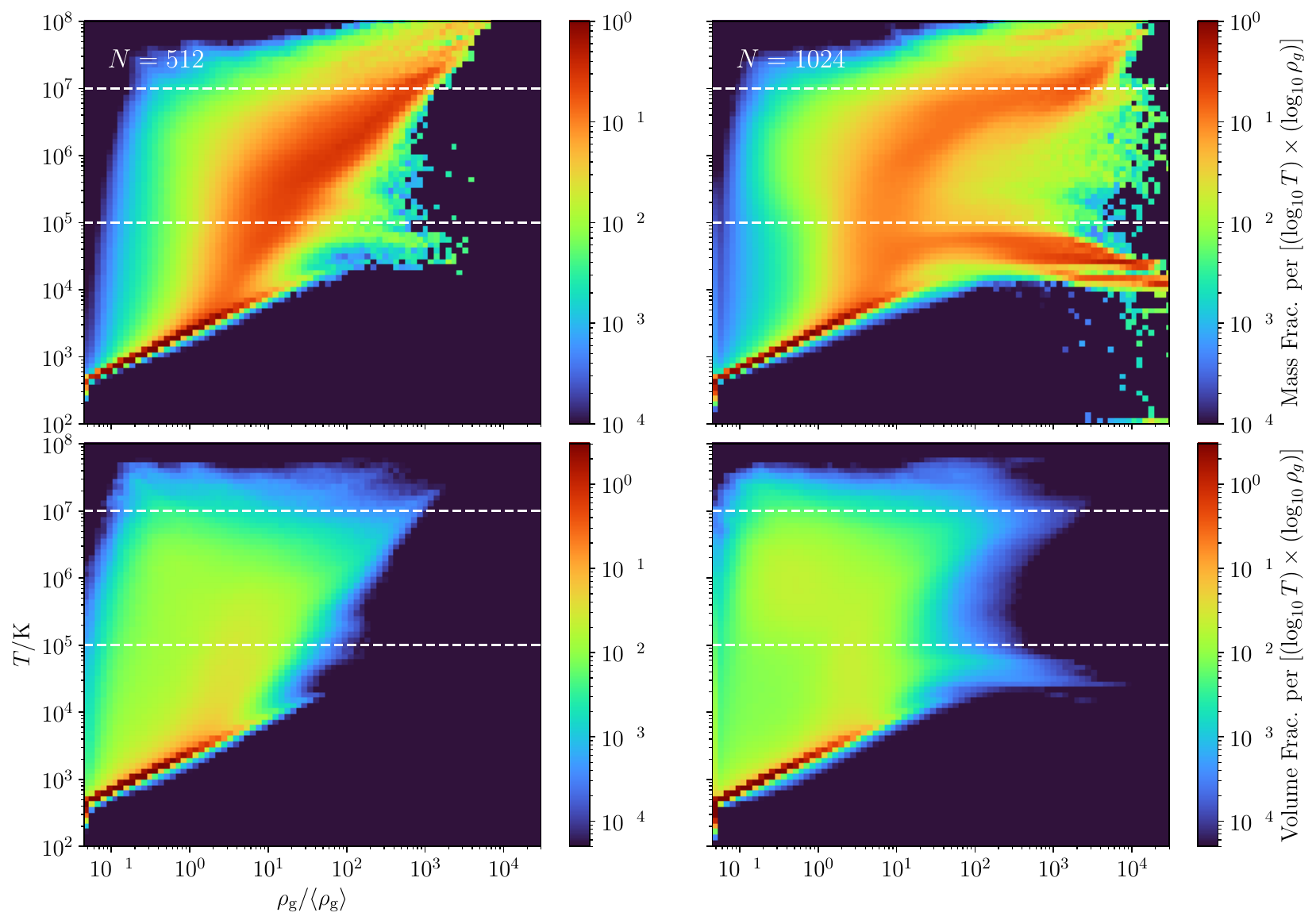}
  \caption{Similar to Figure~\ref{fig:histogram_dual} but
    for $N=512$ (left column) and $N=1024$ (right column)
    simulations.  }
  \label{fig:lowres-hist}
\end{figure*}

One of the key features presented in this study is the
implementation of significantly higher numerical resolution
compared to existing cosmological simulations that
investigate the WHIM. The role of enhanced resolution is
demonstrated in Figure~\ref{fig:lowres-hist}, where
discernible differences emerge in the two-dimensional
histogram within the
$\log_{10}T \times \log_{10}\rho_{\rm g}$ phase
space. Notably, horizontal ``streaks'' become apparent at
gas densities
$\rho_{\rm g} \gtrsim 10^{1.5} \mean{\rho_{\rm g}}$ and
temperatures $T \gtrsim 7\times 10^{4}~\K$ (corresponding to
the prominent cooling features associated with carbon and
oxygen) as well as at $T \sim 10^4~\K$ (corresponding to the
characteristic \lya{} cooling regime). These thermally
distinct features are exclusively resolved in simulations
with grid resolutions of $N \geq 2048$ along each dimension.

To quantitatively assess the resolution requirements, we
compare our simulation cell size
$\Delta x = 36.2~\kpc \times (4096/N)$ against the Jeans
length for gas at temperatures above $\sim 10^5~\K$, given
by:
\begin{equation}
  \label{eq:len-jeans}
  \lambda_{\rm J} \sim 300~\kpc \times
  \left( \dfrac{T}{10^5~\K} \right)^{1/2}
  \left( \dfrac{\rho_{\rm g}}{10^3\mean{\rho_{\rm g}}}
  \right)^{-1/2}.
\end{equation}
The low-resolution simulations depicted in
Figure~\ref{fig:lowres-hist} fail to adequately resolve
structures on the scale of $\lambda_{\rm J}$ with a
sufficient number of computational cells. Specifically, the
$N=512$ model ($\Delta x = 289~\kpc$) is fundamentally
incapable of resolving the Jeans length, while the $N=1024$
run ($\Delta x = 145~\kpc$) still provides insufficient
spatial sampling to accurately capture the dynamics of
self-gravitating collapse processes.

Hereafter spatial scales are quoted as proper (physical)
kiloparsecs; the fiducial $N=4096$ resolution of
$\Delta x = 36.2~\kpc$ proper corresponds to
$\Delta x = 24.5~\kpc~h^{-1}$ comoving for the adopted
Planck cosmology ($h=0.674$). The Jeans-length
normalization adopted in this appendix
($\lambda_{\rm J}\sim 300~\kpc$ at $10^5~\K$,
$10^3\mean{\rho_{\rm g}}$) is equivalent to the body's
formulation (\S\ref{sec:result-obv-lya};
$\lambda_{\rm J}\sim 20~\kpc$ comoving at $10^4~\K$,
$10^{-2}~m_p~\cm^{-3}$) once the
proper-to-comoving conversion and the
$\mean{\rho_{\rm g}}\to\rho_{\rm c1}$ mapping are accounted
for.

\subsection{Comparison with Moving-Mesh and AMR
  Simulations}

This resolution limitation is further contextualized through
comparison with the Illustris-TNG simulation suite
\citep{2018MNRAS.473.4077P, 2018MNRAS.475..648N}. As
Illustris-TNG employs a moving-mesh code, its computational
resources are preferentially allocated to regions within and
around dark matter halos, optimizing for studies of galactic
formation and evolution. For instance, the TNG100 simulation
initializes with a baryonic mass resolution of
$\Delta m_{\rm baryon} = 1.4\times 10^6~M_\odot$ per
  cell \citep{2018MNRAS.473.4077P}. The corresponding
  equivalent spatial resolution can be
estimated as:
\begin{equation}
  \label{eq:res-tng}
  \Delta x \sim 60~\kpc\times
  \left( \dfrac{\rho_{\rm g}}{\mean{\rho_{\rm
          g}}}\right)^{-1/3}    
  \left( \dfrac{\Delta m_{\rm baryon}}{1.4\times
      10^6~M_\odot}\right)^{1/3}. 
\end{equation}
This level of resolution remains inadequate for resolving
the self-gravitating Jeans scale $\lambda_{\rm J}$, as the
mass per cell $\Delta m_{\rm baryon}$ becomes even
significantly lower in the diffuse WHIM
environment. Although the TNG50 simulation achieves higher
mass resolution, its relatively constrained simulation
volume fails to encompass the full dynamic range of scales
that exhibit the highest amplitude in the large-scale
structure power spectrum, thereby limiting its statistical
robustness for WHIM studies.

Adaptive mesh refinement (AMR) codes face analogous
limitations for WHIM studies. Although AMR achieves
extremely high resolution inside collapsed halos, the
diffuse WHIM filaments are typically refined less
aggressively, because the refinement criteria are usually
tied to mass or density thresholds. Moreover, the
non-uniform resolution complicates the interpretation of
statistical quantities such as phase distributions, since
the effective spatial resolution then varies systematically
with environment. The uniform high resolution of the
present simulation avoids this bias, guaranteeing that the
resolved physics does not depend on the local refinement
level.

\subsection{Projected Structure and Synthetic Emission
  Observables}

\begin{figure*}
  \centering
  \includegraphics[width=0.92\textwidth, keepaspectratio]
  {\figdir/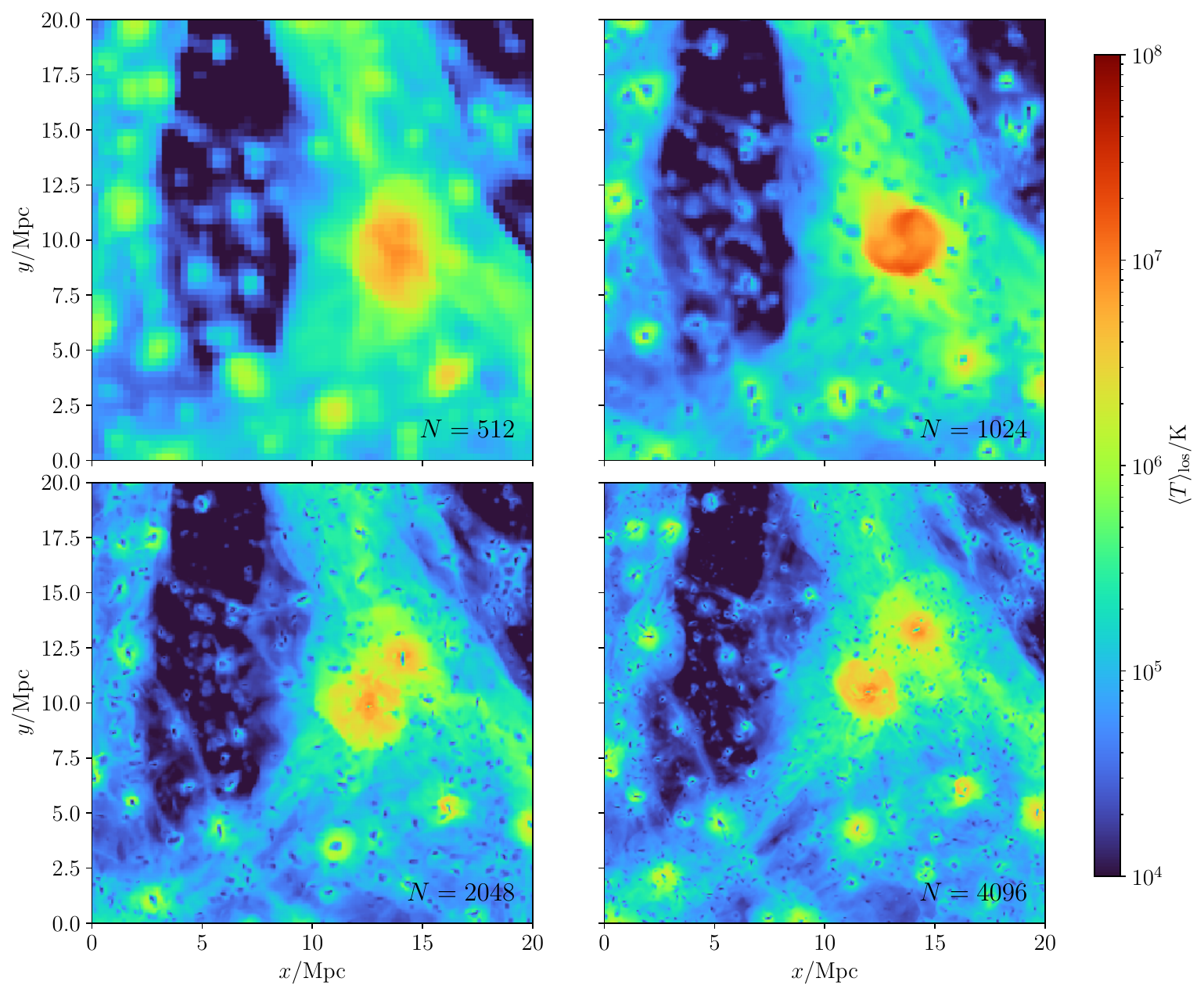}
  \caption{Line-of-sight effective temperature,
    $\langle T\rangle_{\rm los}\equiv(\int p{\rm d}z)
    /(\int\rho_{\rm g}\,{\rm d}z)$, with the quotient
    converted to Kelvin, in a $20\times20$~Mpc zoom-in at
    the lower-left corner of the box at $z\simeq0$, for
    $N=512$ (upper left), 1024 (upper right), 2048 (lower
    left), and 4096 (lower right). Cool ($\sim10^4$~K)
    compact regions that coincide with the overdense knots
    in Figure~\ref{fig:zoom-rho-res} appear for
    $N\geq1024$ but not for $N=512$. }
  \label{fig:zoom-T-res}
\end{figure*}
\begin{figure*}
  \centering
  \includegraphics[width=0.92\textwidth, keepaspectratio]
  {\figdir/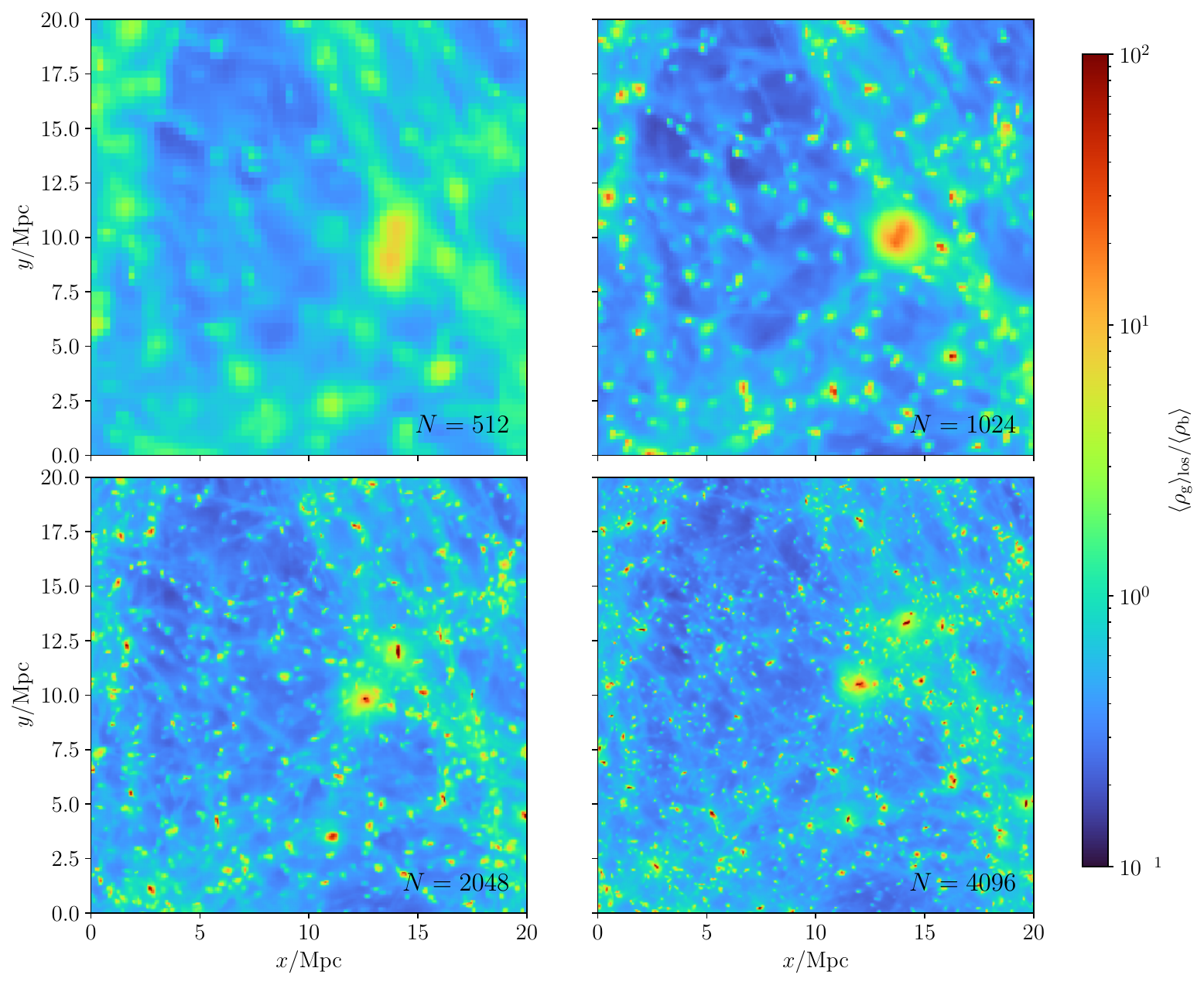}
  \caption{Projected gas density averaged along the line of
    sight, $\langle\rho_{\rm g}\rangle_{\rm los}\equiv
    L_{\rm box}^{-1}\int\rho_{\rm g}\,{\rm d}z$, normalized
    to the mean baryon density $\langle\rho_{\rm b}\rangle$,
    in the same zoom-in region and panel layout as
    Figure~\ref{fig:zoom-T-res}. }
  \label{fig:zoom-rho-res}
\end{figure*}

The implications of numerical resolution extend
directly to the thermodynamic structure of the gas in
projection. Figures~\ref{fig:zoom-T-res}
and~\ref{fig:zoom-rho-res} compare the line-of-sight
effective temperature and the projected gas density in a
$20\times20$~Mpc zoom-in for all four resolutions. For
$N\geq1024$ the temperature maps are dotted with cool
($\sim10^4$~K) compact regions that almost always overlap
with the overdense knots in
$\langle\rho_{\rm g}\rangle_{\rm los}$ --- collapsed gas
that has been able to cool because the resolution is
sufficient to follow the collapse--cooling sequence. No
such cool spots appear at the density peaks of the $N=512$
run: the same structures remain at
$\gtrsim10^5$--$10^6$~K, i.e.\ the gas is artificially held
at WHIM temperatures, consistent with its inflated WHIM
mass fraction (Figure~\ref{fig:frac-whim}). This systematic
effect carries over to synthetic observables: inadequate
resolution cannot properly address relatively high density
regions, while the emissivity per volume is generally
$\proptosim\rho_{\rm g}^2$, so the peak intensities and
integrated surface brightness of emission from halo
interfaces and filamentary structures are substantially
diminished, with negative impact on the interpretability of
mock observations designed to guide future observational
campaigns.

\begin{figure*}
  \centering
  \includegraphics[width=0.92\textwidth, keepaspectratio]
  {\figdir/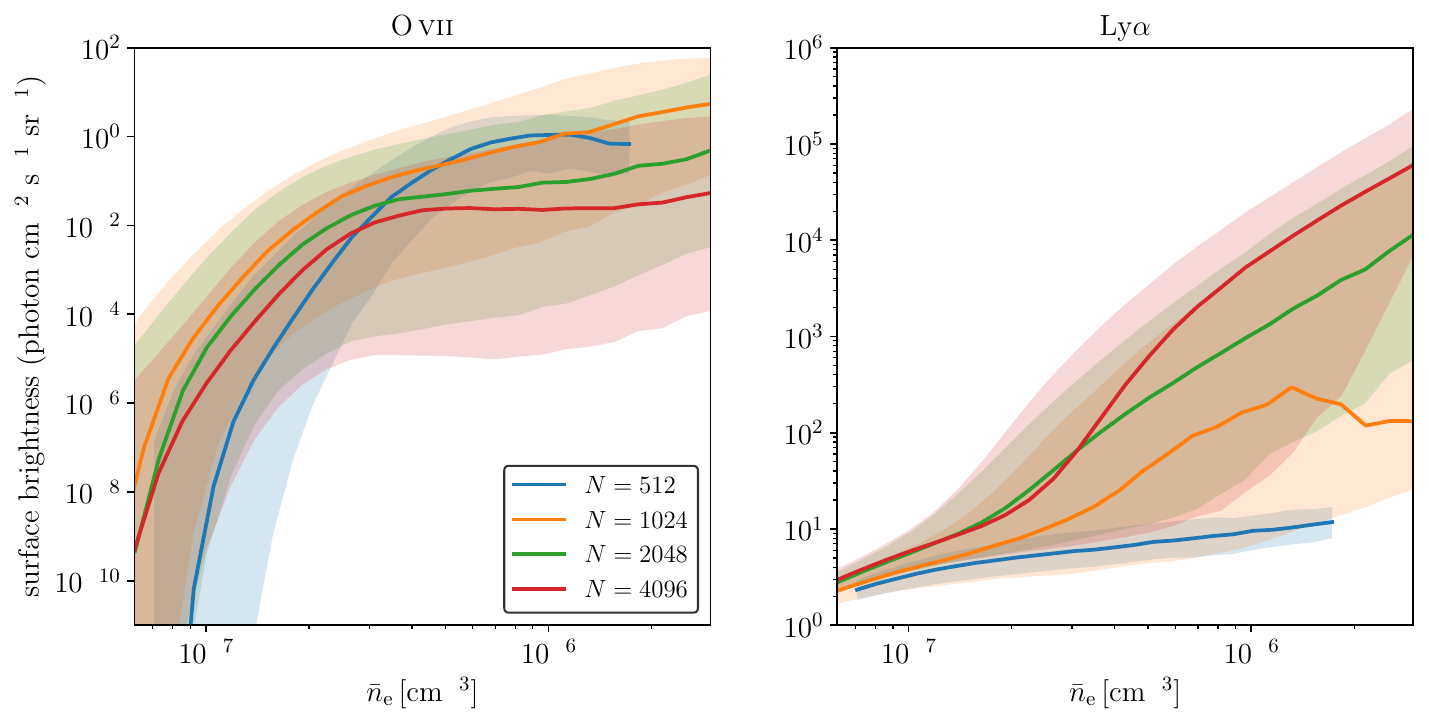}
  \caption{Surface brightness per sightline versus the mean
    free-electron density along the line of sight,
    $\bar n_{\rm e}\equiv L_{\rm box}^{-1}\int n_{\rm e}\,dl$,
    at $z\simeq 0$: \ion{O}{vii} (left panel) and \lya{}
    (right panel). Solid lines are binned medians and the
    shaded bands enclose the 16th--84th percentiles, for
    $N=512$ (blue), 1024 (orange), 2048 (green), and 4096
    (red). The \lya{} brightness, weighted by the squared
    gas density, keeps rising with resolution at fixed
    $\bar n_{\rm e}\gtrsim 10^{-6}~\cm^{-3}$ as increasingly
    dense substructure is resolved, whereas the
    temperature-selective \ion{O}{vii} emission changes by
    progressively smaller factors and declines at high
    $\bar n_{\rm e}$ once cold dense cores are resolved out
     of the \ion{O}{vii} temperature window. The \lya{}
     surface brightness rises by a factor of $\sim 5$ per
     doubling at $N\geq 2048$ and has not yet converged at
     $N=4096$, because it scales as $n_e^2$ and is dominated
     by unresolved dense substructure.}
  \label{fig:emis-ne-res}
\end{figure*}

This density-squared scaling is quantified directly in
Figure~\ref{fig:emis-ne-res}, which compares the
per-sightline surface brightness of \ion{O}{vii} and \lya{}
against the mean electron density along the line of sight
for all four resolutions. At fixed mean density, the median
\lya{} brightness keeps rising with resolution---by a factor
of $\sim 5$ between $N=2048$ and $N=4096$ alone at
$\bar n_{\rm e}\sim 10^{-6}~\cm^{-3}$, with no sign of
convergence---because the emission is dominated by the
densest substructure along the line of sight, which coarser
grids cannot resolve. The \ion{O}{vii} relation, weighted
additionally by its temperature window, behaves more
favorably: the change per resolution doubling shrinks to a
factor of a few between $N=2048$ and $N=4096$, and the
median brightness even declines at the highest densities,
where resolving cold dense cores removes gas from the
\ion{O}{vii}-emitting phase. Density-squared-weighted
observables are therefore not yet converged even at
$N=4096$, whereas temperature-selective tracers such as
\ion{O}{vii} are substantially more robust.

\end{document}